\documentclass[a4paper,11pt]{article}
\usepackage{jheppub}
\usepackage{subfigure}
\usepackage{relsize}

\def  \bcen   {\begin{center}}
\def  \ecen   {\end{center}}
\def  \beq    {\begin{equation}}
\def  \eeq    {\end{equation}}
\def  \beqa   {\begin{eqnarray}}
\def  \eeqa   {\end{eqnarray}}

\def\bea{\begin{eqnarray}}
\def\eea{\end{eqnarray}}

\title{Vector like leptons with extended Higgs sector }

\author[a]{Sumit K. Garg,}
\author[a]{C.S. Kim,}

\emailAdd{sumit@yonsei.ac.kr}
\emailAdd{cskim@yonsei.ac.kr}

\abstract{We examined the influence of additional scalar doublet on the parameter space of the Standard Model  supplemented with a generation of
new vector like leptons. In particular we identified the viable regions of parameter space by inspecting various  constraints especially electroweak precision (S, T and U)
parameters. We  demonstrated that the additional scalar
assists in alleviating the tension of electroweak precision constraints  and thus permitting
larger Yukawa mixing and mass splittings among vector like species. We also compared and  contrasted the regions of parameter space pertaining
to the latest LHC Higgs to diphoton channel results in this scenario with vector like leptons in single Higgs doublet and pure two Higgs doublet model  case.}


\begin{document}
\maketitle
\section{Introduction}

ATLAS \cite{ATLAS:2012ae} and CMS \cite{Chatrchyan:2012tx} collaborations at the CERN Large Hadron Collider (LHC) reported the
observation of much awaited scalar resonance with  mass of discovered field hovering around $\sim$ 126\,GeV. Now with analysis of around 25\,$fb^{-1}$
of collider data, the LHC almost confirmed \cite{MoroindHiggs} the new state to be a Higgs boson.
However, still much remains to be seen when the experimentally observed properties of new
state will confront with the predictions of  highly
celebrated Standard Model (SM).
Thus any reported deviation from its expected SM properties could
signal the presence of physics beyond the SM (BSM).

Among innumerable extensions of the SM that can modify Higgs physics, supplementing the SM with additional fermions serves as one of the phenomenologically
interesting and much investigated scenarios of physics BSM.
Here new fermionic fields by circulating in loops related to production and decay of Higgs can significantly
alter the predictions of the SM. Much studied cases of the SM with chiral 4th generation \cite{chiral4G}, vector like fermions \cite{VecEwpr,vecferms}
etc. come under these category of models. These kind
of scenarios are also studied extensively within extended Higgs sector like two Higgs doublet model (2HDM) \cite{4G2HDM}, which not only
provides rich phenomenology testable at the LHC
but also provide implications which are completely different from single Higgs case.

Motivated from these observations we will investigate here one such possibility. We will take up  the  case of
vector like leptons \cite{vecferm1,vecferm2,vecferm3,vecferm4,vecferm5}  in 2HDM scenario. The similar studies for supersymmetric case are abundantly discussed \cite{susyvecferm} in
literature.
The recent implications of these scenarios are also studied \cite{susyvecfermRec} with much interest. Since additional states do not carry any strong color charge
so they will only contribute into
the decay loops of Higgs and thus can act as much sensitive probe of new physics. Here for simplicity we will introduce one complete generation of vector like leptons
in addition to the SM fermionic fields. Regarding scalar sector, various
versions of 2HDM were introduced in literature depending on
the coupling of Higgs to fermions. Since newly discovered resonance looks more like the SM state so we will consider
the simplest case of inert doublet model \cite{idmall}. In this model there is no mixing between two doublets, and thus the lightest CP even state plays the role of
the SM Higgs.

The new states will  contribute in self energy diagrams of electroweak gauge bosons and thus are constrained from electroweak precision
observables. These effects on electroweak precision parameters can be parametrized by three gauge self-energy
parameters (S, T, U) introduced by Peskin and Takeuchi \cite{PeskinSTU}. As we will discuss in
numerical section  with additional doublet it is possible to have cancellations between scalar and fermionic contributions and thus alleviating these constraints. This will
in turn permit larger Yukawa mixing and mass splittings among vector like states. Apart from this the parameter space is also constrained by
theoretical constraints
like vacuum stability, perturbativity and unitarity which come into picture due to extended scalar sector.

Among  various properties of newly discovered scalar field, its loop induced decays can serve as a much sensitive probe of new physics. Here BSM fields
that couple to Higgs can challenge the expectations of the SM by circulating in its loop decay diagrams.
In this scenario, the charged fermionic and scalar fields contribute in the loop induced decays of Higgs and thus can give the signatures
which are completely different from single Higgs doublet case. In particular we have chosen the process Higgs to gamma gamma as a signature of BSM physics.
ATLAS and CMS reported an excess in this channel \cite{ATLAShtogmgm, CMShtogmgm} with signal strength around 1.5.
However, these results were updated  at Moriond Conference by analyzing around 25 $fb^{-1}$ of collider data. ATLAS reported an excess \cite{ATLAShtogmgmnew} in this channel
with signal
strength $\sigma/\sigma_{SM} = 1.65 \pm 0.24(stat)^{+0.25}_{-0.18}(syst)$ while CMS number comes down \cite{CMShtogmgmnew} from $\sigma/\sigma_{SM}= 1.56 \pm 0.43$ to
$\sigma/\sigma_{SM}= 1.11 \pm 0.31$ with cut based events, and $0.78\pm 0.27$ with selected and categorized events. Thus results in this channel are not
entirely consistent with the SM and need further data analysis. In this work we investigated the situation of Higgs to gamma gamma in allowed parameter space and contrasted
it with  vector like leptons in single Higgs doublet and  2HDM case.
A  complete analysis for all the channels in this kind of scenario is deferred for future publication.

The paper is organized as follows. In the next section we will introduce the general setup of the model including
field content and various interactions. Electroweak precision tests and other theoretical constraints are explained
in Sec.\,3 while Higgs to gamma gamma decay rate is analyzed in Sec.\,4. In Sec.\,5 we present the numerical results of our
study and finally we conclude with summary and discussion of our results in Sec.\,6. Main formulae related to this
study are presented in the appendix.

\section{General Setup}

In addition to the SM fermionic and gauge fields we have two scalar doublets ($\Phi$, $\Phi_I$) and vector
like leptons which have following transformation properties under $SU(3)_C \times SU(2)_L \times U(1)_Y$:
\begin{eqnarray}
&&\Phi = ({\bf 1}, {\bf 2}, 1/2),
\hspace{6mm}
\Phi_I = ({\bf 1}, {\bf 2}, 1/2),
\hspace{0.7cm}
L_4 = ({\bf 1}, {\bf 2}, -1/2),\nonumber\\
&&
L_4^c = ({\bf 1}, {\bf 2}, -1/2),
\hspace{1cm}
E_4^c = ({\bf 1}, {\bf 1}, -1),
\hspace{1cm}
E_4 = ({\bf 1}, {\bf 1}, -1).
\end{eqnarray}

Here we consider the scenario  where only  doublet $\Phi$, not $\Phi_I$, couples to all fermions
and vector like species while its counterpart
remains inert w.r.t  Yukawa interactions.  this inert doublet model \cite{idmall} does have gauge interactions. This task
can be achieved by assuming $Z_2$ symmetry under which $\Phi_I$ is odd while $\Phi$ is even.

The scalar  potential that
can be formed using these two Higgs doublets is given by \cite{Gunionetal2HDM,hgmgmidm2}
\begin{equation}
  \begin{aligned}
    V = &\mu^2\Phi^\dagger\Phi +\mu_{I}^2\Phi_I^\dagger\Phi_I
    +\frac{1}{2}\lambda_1\left(\Phi^\dagger\Phi\right)^2
    +\frac{1}{2}\lambda_2\left(\Phi_I^\dagger\Phi_I\right)^2
    \\
    &
    +\lambda_3\left(\Phi^\dagger\Phi\right)\left(\Phi_I^\dagger\Phi_I\right)
    +\lambda_4\left(\Phi^\dagger\Phi_I\right)\left(\Phi_I^\dagger\Phi\right)
    +\left\{
    \frac{1}{2}\lambda_5\left(\Phi^\dagger\Phi_I\right)^2
        +\mathrm{h.c.}\right\}.
  \end{aligned}
  \label{eq:pot_gen}
\end{equation}
The part describing the interaction of new vector like leptons is given by \cite{vecferm1,vecferm2}
\[ -{\cal{L}}\hspace{2mm} =\hspace{2mm} y_4 \bar{E_4^c} L_4 \tilde{\Phi} \hspace{2mm} + y_4 \bar{L^c_4} E_4 \Phi \hspace{2mm} + m_{L_4} \bar{L^c_4}  L_4 \hspace{2mm}
+ m_{E_4} \bar{E^c_4}  E_4 \hspace{2mm} + h.c. \]

where $\tilde{\Phi}= i \tau_2 \Phi^* $, $y_4$ is the yukawa coupling to the Higgs and $\{ m_{L_4}, m_{E_4} \}$ are vector like mass parameters. Here
we neglect any mixing between the SM fermions and new vector like leptons in order to avoid any additional constraints from
flavor violation. The electroweak gauge symmetry is broken when Higgs field attains a vacuum expectation value and thus the scalar
doublets can be expressed in terms of physical fields as
\beq
 \Phi  = \left( \begin{array}{c} \phi^+ \\ \frac{v+h + i \chi}{\sqrt{2}} \end{array}
\right) ~~~~,~~
 \Phi_I  = \left( \begin{array}{c} \phi_I^+ \\ \frac{(S+ i A)}{\sqrt{2}} \end{array}
\right)~.
\eeq
Thus the scalar spectrum of this model consists of
two CP even neutral scalars ($h$, $S$) one CP odd neutral
scalar ($A$) along with  a pair of charged scalars
($H^\pm$). Here $h$ plays the role of the SM Higgs Boson.

The fermion mass matrix for vector like states becomes of the form
\begin{equation}
{\cal{M}} = ( \bar{E_4^c} \hspace{2mm} \bar{L_4^c}^+) \left( \begin{array}{cc}
m_{E_4} & \frac{y_4 v}{\sqrt{2}}  \\
 \frac{y_4 v}{\sqrt{2}} &  m_{L_4}
\end{array}
\right)
\left( \begin{array}{c}
E_4\\
L_4^-
 \end{array}\right)\, .
\label{eq:mass-2+1}
\end{equation}
The mass matrix ${\cal{M}}$ can be diagonalized by the transformation
${\cal{M}}_D = U_L {\cal{M}} U_R^{\dag}$ with the eigenstates
\begin{equation}
l \equiv
\left( \begin{array}{c}
l_{1}\\
l_{2}
 \end{array}\right) = U_{R}
\left( \begin{array}{c}
E_4\\
L_4^{-}
 \end{array}\right)~~ {\rm and}~~
\left( \begin{array}{c}
\bar{l}_{1}\\
\bar{l}_{2}
 \end{array}\right) = U_{L}^{*}
\left( \begin{array}{c}
\bar{E_4^c}\\
\bar{L_4^c}^{+}
 \end{array}\right)~,
\label{eq:matdef}
\end{equation}
where $U_{L,R}$ are unitary matrices. Thus spectrum consists of
mass eigenstates $l_1$, $l_2$ in charged sector  and
$N_4$ in neutral sector with masses
\begin{equation}
M_{l_1,l_2} = \frac{1}{2}\left[ (m_{L_4} + m_{E_4}) \mp \sqrt{(m_{L_4}-m_{E_4})^2 + 2 y_4^2 v^2}\right]
\quad {\rm and } \quad
M_{N_4} = m_{L_4}.
\label{m1m2-21}
\end{equation}

\section{Constraints on Parameter Space}

The parameter space of theory can be constrained by enforcing various theoretical and experimental
constraints. We imposed the following restrictions in our parameter scan.

\subsection{Perturbativity}

We demand perturbativity of all quartic couplings by imposing condition
\begin{center}
$|\lambda_i| \leq 4 \pi $~.
\end{center}

\subsection{Vacuum Stability}

The requirement of positivity of the potential enforces the following conditions on the quartic
couplings \cite{2HDMsymbrk}
\begin{eqnarray}
\lambda_{1,2} > 0 \quad \rm{and} \quad \lambda_3 + \lambda_4 -|\lambda_5| +
\sqrt{\lambda_1 \lambda_2} >0 \quad\rm{and} \quad\lambda_3+ \sqrt{\lambda_1
  \lambda_2} > 0~.
\end{eqnarray}

\subsection{Unitarity}

Here one can obtain constraints on model parameters by requiring the tree level unitarity for
the scattering of Higgs bosons and longitudinal parts of the EW gauge bosons \cite{unitarity1,unitarity2}. In 2HDM the necessary
and sufficient conditions for the S-matrix to be unitarity in terms of its eigenvalues are derived
in \cite{unitarity2HDM}. The eigenvalues of S-matrix are given by
%
\begin{eqnarray}
&&L_{1,2}=\lambda_3 \pm \lambda_4 \quad , \quad
L_{3,4}= \lambda_3 \pm \lambda_5~,\\
&&L_{5,6}= \lambda_3+ 2 \lambda_4 \pm 3\lambda_5\quad , \quad
L_{7,8}=\frac{1}{2}\{ \lambda_1 + \lambda_2 \pm \sqrt{(\lambda_1 - \lambda_2)^2 + 4 \lambda_4^2} \}~,
\\
&&
L_{9,10}= \frac{1}{2}\{ 3\lambda_1 + 3\lambda_2 \pm \sqrt{9(\lambda_1 - \lambda_2)^2 + 4 (2\lambda_3 +
   \lambda_4)^2} \}~,
\\
&&
L_{11,12}=
 \frac{1}{2} \{ \lambda_1 + \lambda_2 \pm \sqrt{(\lambda_1 - \lambda_2)^2 + 4 \lambda_5^2} \}~.
\end{eqnarray}
We impose perturbative unitarity constraint on all eigenvalues $L_i$'s by requiring:
\beq
|L_i| \le 8 \pi ~, \forall ~ i = 1,...,12.
\eeq

\subsection{Electroweak Precision Constraints}

Apart from setting the direct detection limits on new physics models, physics beyond the SM can  also be constrained through
its effect on electroweak precision observables. In other words any new physics model should confront with the tremendous success
of the SM. The new physics fields mark their presence through the contribution to the vacuum polarization
diagrams of the electroweak gauge bosons ~\cite{Altarelli:1990zd,PeskinSTU}. These effects on electroweak precision parameters
can be parametrized by three gauge self-energy
parameters (S, T, U) introduced by Peskin and Takeuchi \cite{PeskinSTU}:
\begin{eqnarray}
\alpha(M_Z^2) \, S^{\text{NP}} & = & \frac{4 s_W^2 c_W^2}{M_Z^2} \left [\Pi^{\text{NP}}_{ZZ} (M_Z^2) - \Pi^{\text{NP}}_{ZZ} (0)
    -\Pi^{\text{NP}}_{\gamma \gamma}(M_Z^2) - \frac{c_W^2-s_W^2}{c_W
        s_W} \, \Pi^{\text{NP}}_{\gamma Z}(M_Z^2)\right] ,\nonumber
      \\
\alpha(M_Z^2) \, T^{\text{NP}} & = & \frac{\Pi^{\text{NP}}_{WW}(0)}{M_W^2}
        - \frac{\Pi^{\text{NP}}_{ZZ}(0)}{M_Z^2} ,\nonumber
          \\
\alpha(M_Z^2) \, U^{\text{NP}} & = & 4 s_W^2 \left [
            \frac{\Pi^{\text{NP}}_{WW}(M_W^2)-\Pi^{\text{NP}}_{WW}(0)}
                {M_W^2} - c_W^2 \left(
                \frac{\Pi^{\text{NP}}_{ZZ}(M_Z^2)-\Pi^{\text{NP}}_{ZZ}(0)}{M_Z^2}\right)
                    \nonumber \right .\\ & & \left . - 2 s_W c_W \,
                    \frac{\Pi^{\text{NP}}_{\gamma Z}(M_Z^2)}{M_Z^2} -
                       s_W^2 \, \frac{\Pi^{\text{NP}}_{\gamma
                            \gamma}(M_Z^2)}{M_Z^2} \right ],
\label{eq:STU}
\end{eqnarray}
where $s_W$ is the sine of weak mixing angle $\theta_W$,
$M_Z$ and $M_W$ are, respectively, the Z boson and W boson masses.
For a reference Higgs mass of  $m_{h,\rm ref} = 126$\,GeV and a top quark mass of
$m_{t,\rm ref}= 173$\,GeV, and the following fitted values are obtained~\cite{STUcurrent}
when compared with theory predictions
\begin{eqnarray}
\Delta S & = & S - S_{\rm SM}  =  0.03 \pm 0.10 ,\nonumber \\
\Delta T & = & T - T_{\rm SM}  =  0.05 \pm 0.12 ,\nonumber \\
\Delta U & = & U - U_{\rm SM}  =  0.03 \pm 0.10 ,
\label{eq:stulimits}
\end{eqnarray}
with the associated correlation matrix
\begin{equation}
V= \left( \begin{array}{ccc}
1 & + 0.891 & -0.540 \\
+0.891 & 1 & -0.803 \\
-0.540 & -0.803 & 1
\end{array}\right).
\label{eq:corr}
\end{equation}


In our study we will confront these constraints to our model parameter space by minimizing the $\chi^2_{STU}$
function which is defined as
\begin{equation}
\chi^2_{STU} = \sum_{i,j} (O_i^{\text{NP}} -
O_i)(\sigma^2)^{-1}_{ij}(O_j^{\text{NP}} - O_j),
\label{eq:chi2}
\end{equation}
where $O_i= \Delta S, \Delta T, \Delta U$, are the fitted values of
the oblique parameters with their corresponding uncertainties
$\sigma_i$ defined in Eq. (\ref{eq:stulimits}), $O_i^{\text{NP}} =
S^{\text{NP}}, T^{\text{NP}}, U^{\text{NP}}$ are the contributions
of new physics states and $\sigma^2_{i,j} \equiv \sigma_i V_{ij}\sigma_j$.  Here
$\Delta \chi^2_{STU} = (3.53, 7.81, 11.3)$  correspond to  the (68\%,
95\%, 99\%) Confidence Limit (CL) in a three-parameter fit.

\section{ Higgs to gamma gamma ($h\rightarrow \gamma\gamma$)}
The new charged fields ($l_1,l_2$,$H^\pm$)  make contribution to Higgs
decay width at one loop level.
Thus  Higgs to diphoton decay can be written in terms of the couplings
to the particles in the loop as \cite{vecferm3,hgmgmidm1,hgmgmidm2,hgmgmidm3,hgmgmidm4}

\begin{equation}
\Gamma (h \to \gamma \gamma) = \displaystyle \frac{\alpha^2\,
  m_h^3}{1024 \pi^3} \left \vert \frac{2 }{v}V_{1} (x_W) + \frac{8}{3v}
F_{1/2}(x_t) + \mathlarger{\mathlarger{‎‎\sum}}_{i=1}^{2‎} \frac{\sqrt{2} y_{h f_i \bar f_i}}{m_{f_i}}  F_{1/2}(x_{f_i}) + \frac{C_{hH^{+}H^{-}}}{m_{H^{+}}^2}
C_{0}(x_{H^+}) \right \vert^2  \, ,
\label{eq:hgg}
\end{equation}
where  $x_j \equiv (2 m_j/ m_h)^2$, $j=W,t,f,H^+$, $m_h$ is the Higgs
mass which is fixed to be $126$\,GeV in our study, and $y_{hf\bar{f}}$ ($C_{hH^{+}H^{-}}$) are
the coupling of Higgs to vector like fermions (charged Higgs) with
mass $m_f$ ($m_{H^+}$), respectively. The loop functions $V_{1},
F_{1/2}$ and $C_0$ are defined in the Appendix~\ref{loopfuncts}.

The major loop contributions in the SM come from the top quark and W gauge boson  with a loop
factor  of
$V_1(x_W) \to -8.3$ and  $F_{1/2}(x_t) \to
+1.8$ for $m_h=126$\,GeV. Thus to enhance/suppress $h\to \gamma \gamma$ decay width one requires
constructive/destructive interference between the dominant W boson and new physics sector. Here scalar and fermionic contributions
can also nullify the effect of each other and thus making the decay rate to be completely consistent with the SM value.

Since new states don't contribute in Higgs production channel so we can define following
ratio of decay width which can point out the enhancement/suppression in $h\rightarrow \gamma\gamma$ channel
\begin{equation}
R_{\gamma\gamma} = \frac{\sigma(pp \rightarrow h)}{\sigma_{SM}(pp \rightarrow h)}\frac{\Gamma (h \to \gamma \gamma)}{\Gamma_{SM} (h \to \gamma \gamma)}
=\frac{\Gamma (h \to \gamma \gamma)}{\Gamma_{SM} (h \to \gamma \gamma)}~.
\end{equation}
Thus $R_{\gamma\gamma} < 1$ will correspond to suppression while  $R_{\gamma\gamma} > 1.0 $ enhancement in this channel.

\section{Numerical analysis}

In this section we will discuss the numerical results of our study. First we divulge the constraints from electroweak precision
tests and then identify various parameter regions pertaining to different values of  $R_{\gamma\gamma}$. All other previously mentioned
theoretical constraints are already
included while scanning for the viable model parameter space. The random number generator for the scanning subroutine is taken from the publicly
available code SUSEFLAV\cite{SUSEFLAV}. Since it is possible to express quartic couplings $\lambda_i$ in terms of physical scalar masses
and $\mu_{12}$\cite{hgmgmidm3} so we taken $\{m_A, m_{H^\pm}, m_S\}$ and $\{\lambda_2, \mu_{12}\}$ as our independent parameters. We varied our model parameters in the following 
ranges: \{$m_{L_4}, m_{E_4}\} \in$ [70, 800]\,GeV, $y_4 \in$ [0, 2],
$\lambda_{2} \in$ [0, 4$\pi$], Higgs scalar masses\cite{hgmgmidm3} \{$m_A, m_{H^\pm}, m_S\}\in$ [70, 800]\,GeV and $\mu_{12} \in$ [-500, 500]\,GeV. We also
imposed a constraint of $M_{l_1} > 80$\,GeV to satisfy the direct limit constraints\cite{vecferm3} from the LEP on charged vector like fermions.

\begin{figure}[!t]\centering
\vspace*{-1cm}
\begin{tabular}{c c}
\includegraphics[angle=0,width=75mm]{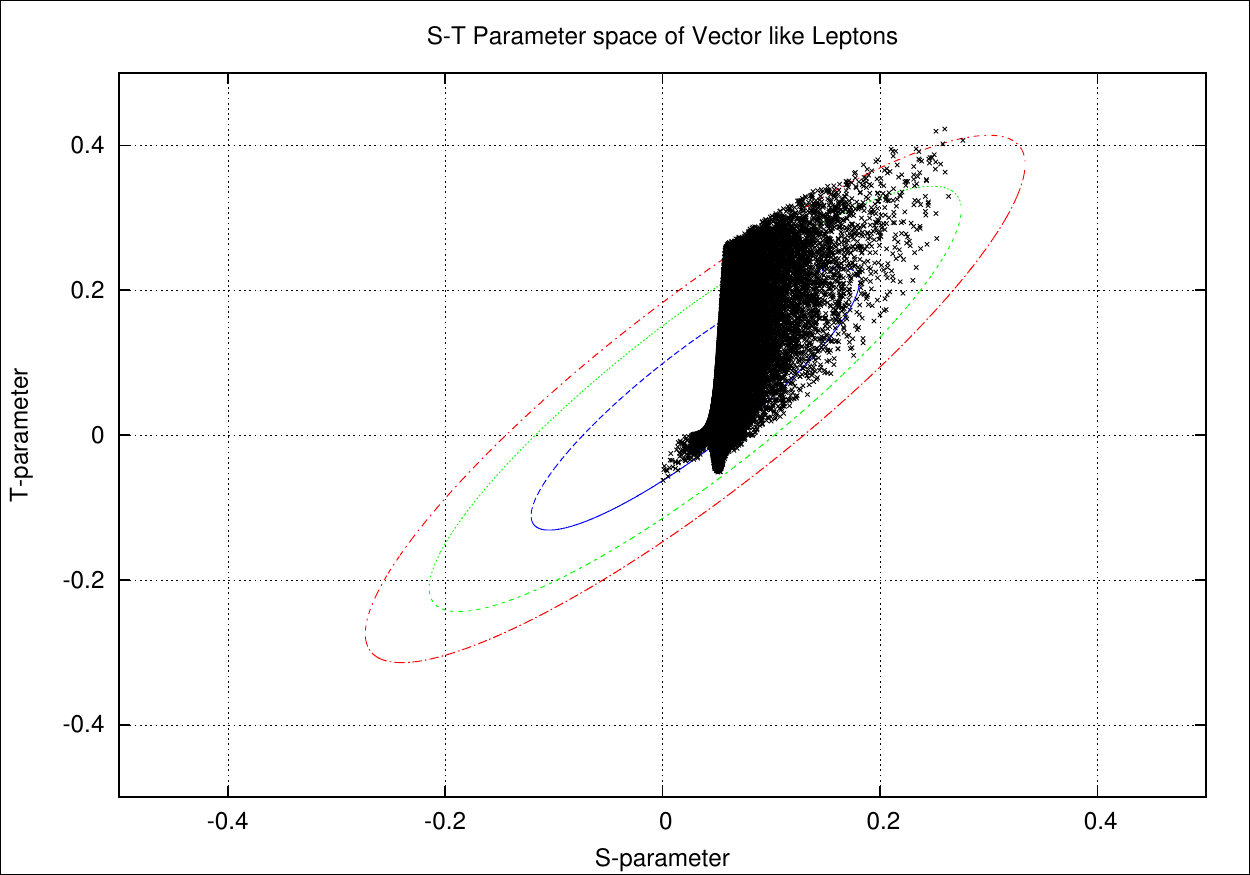} &
\includegraphics[angle=0,width=75mm]{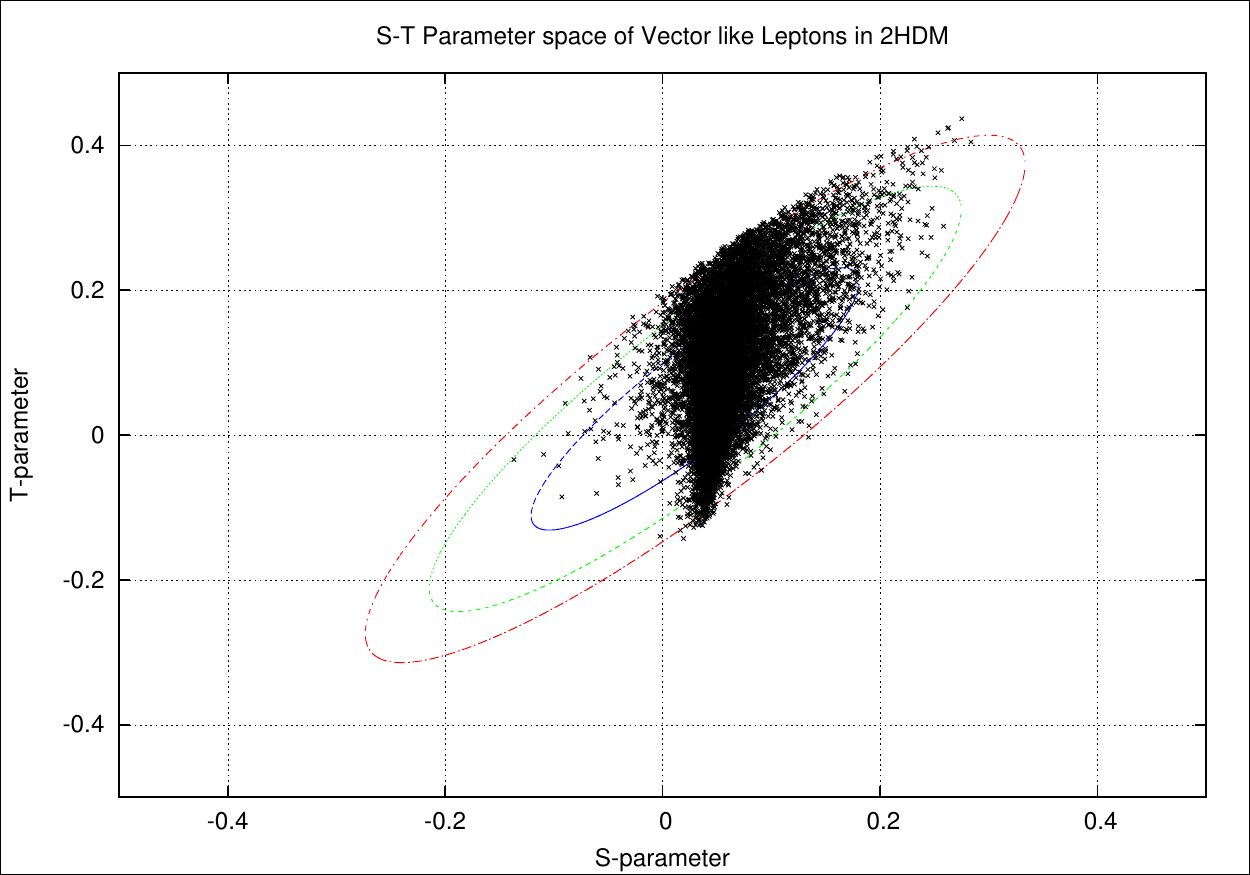}
\end{tabular}
\vskip 1mm
\caption{ S-T scatter plot for vector like leptons with single doublet (left Fig.) and within two Higgs doublet (right Fig.) case. The blue, green and red curves
correspond to $68$\%, $95$\% and $99$\% CL contours.
}
\label{fig.ST}
\end{figure}

In Fig.~\ref{fig.ST}\footnote{SKG would like to thank Daisuke Harada for help in this figure.}  we presented the scatter plot in S-T parameter space for pure 
vector like (left Fig.)
and the case with additional scalar (right Fig.), respectively.
The blue, green and red curves correspond to $68$\%, $95$\% and $99$\% CL contours. Similar plots for U parameter are not
presented here as this parameter doesn't
impose any additional constraint and thus generally neglected in new physics scenarios. As evident from the figures, the pure vector like
case prefers
positive values of S and T  while the contribution of additional scalar shifts the overall region towards central point of the contours and
thus making it
far easier to satisfy electroweak precision data (EWPD) constraints in this case. This is happening due to the cancellations among
scalar and fermionic contributions of these parameters.

\begin{figure}[!t]\centering
\begin{tabular}{c c}
\includegraphics[angle=0,width=75mm]{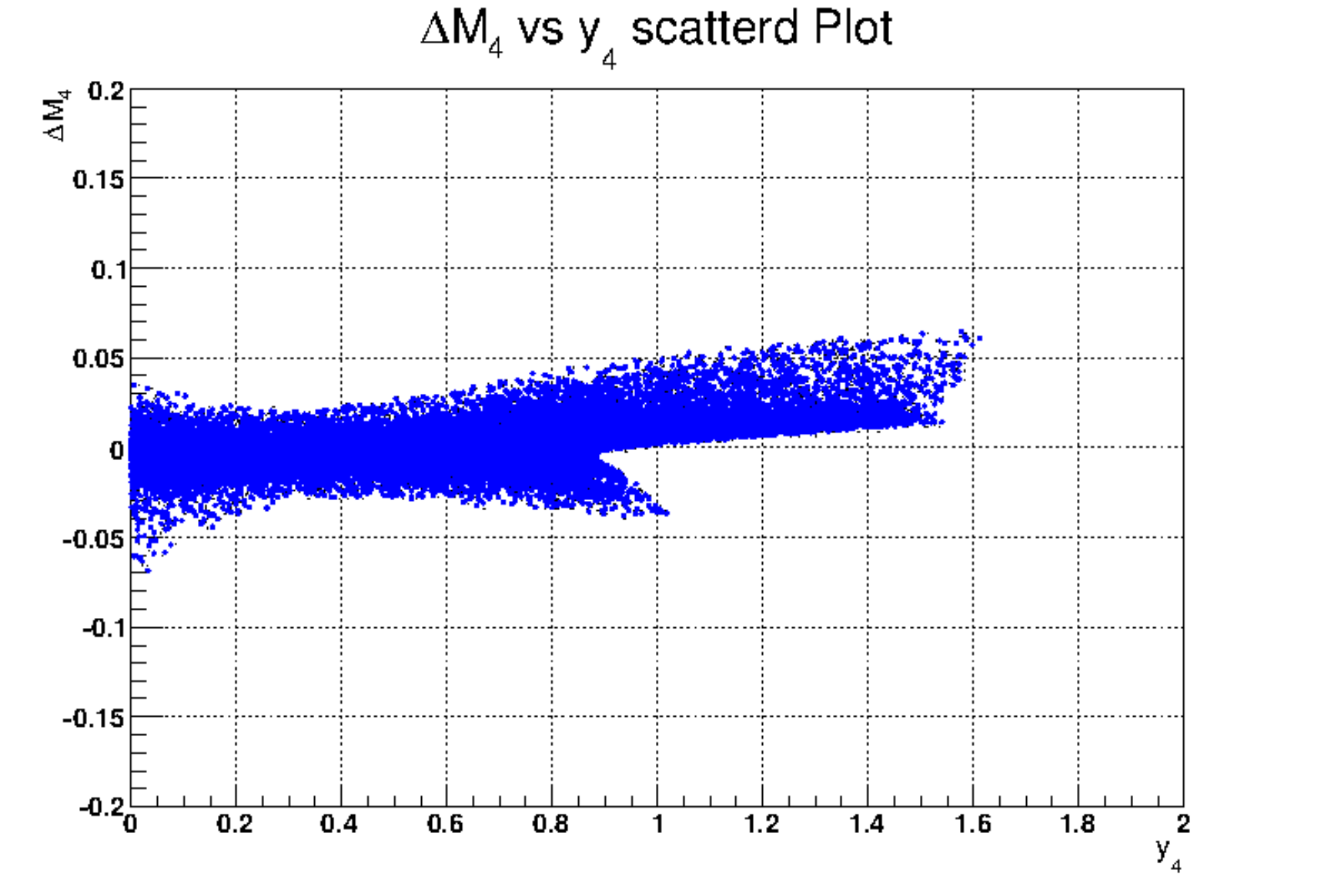} &
\includegraphics[angle=0,width=75mm]{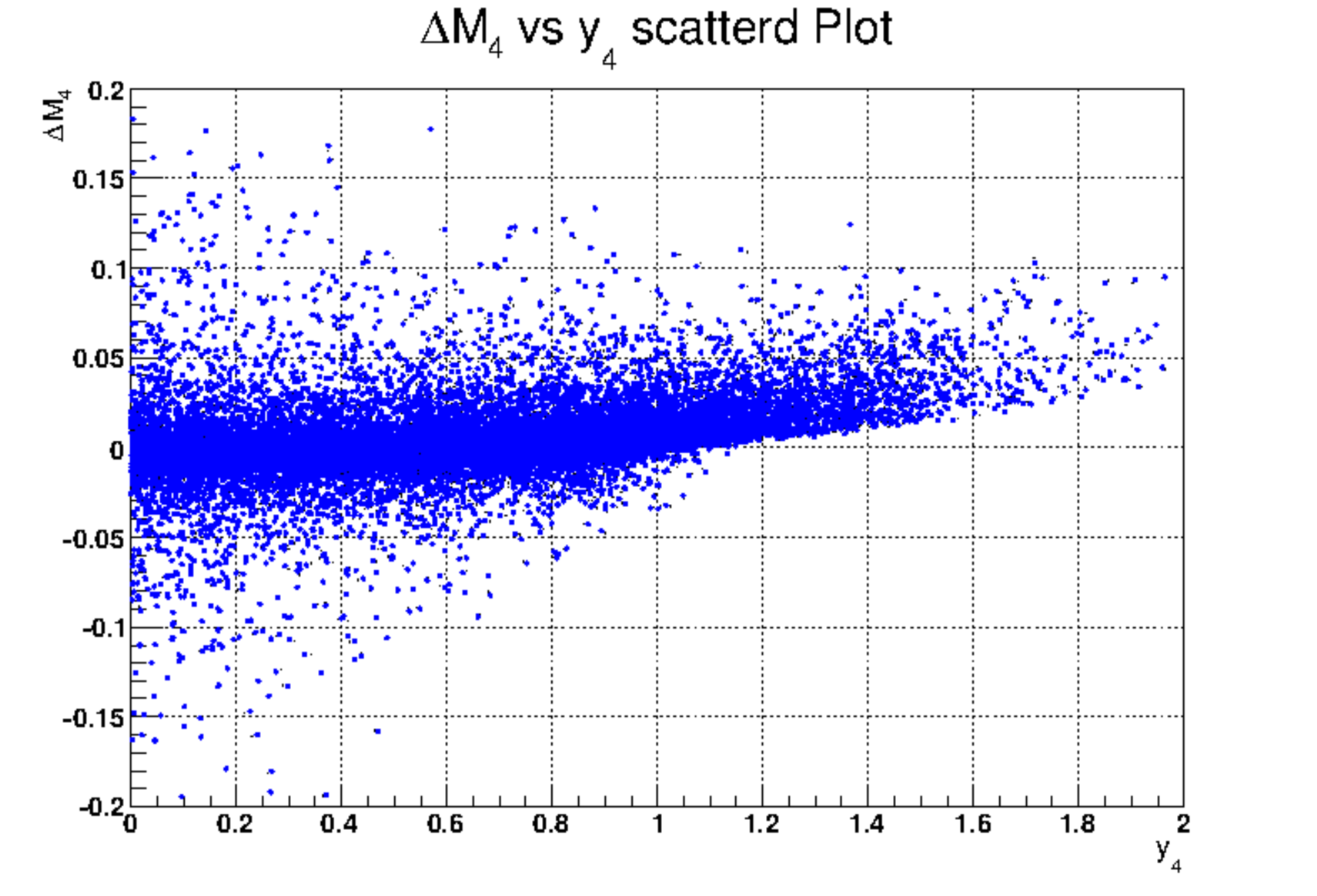}
\end{tabular}
\vskip 1mm
\caption{ Scatter plot of allowed region over $\Delta M_4
(=(m_{L_4} - m_{E_4})/(m_{L_4} + m_{E_4}))$ vs $y_4$ plane  for vector like leptons with single doublet (left Fig.) and within two Higgs doublet (right Fig.) case.
}
\label{fig.splitting}
\end{figure}

\begin{figure}[!t]\centering
\begin{tabular}{c c}
\includegraphics[angle=0,width=75mm]{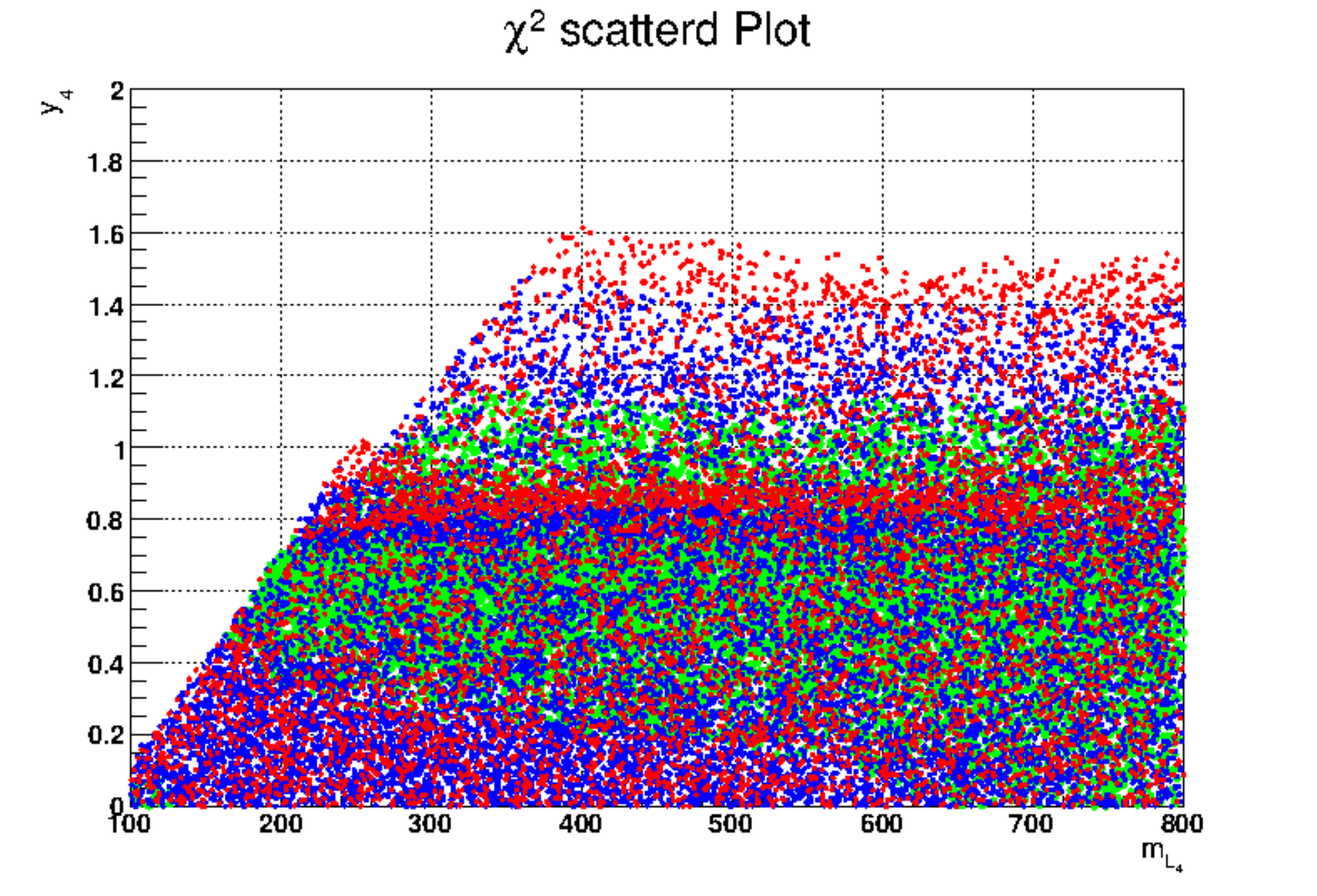} &
\includegraphics[angle=0,width=75mm]{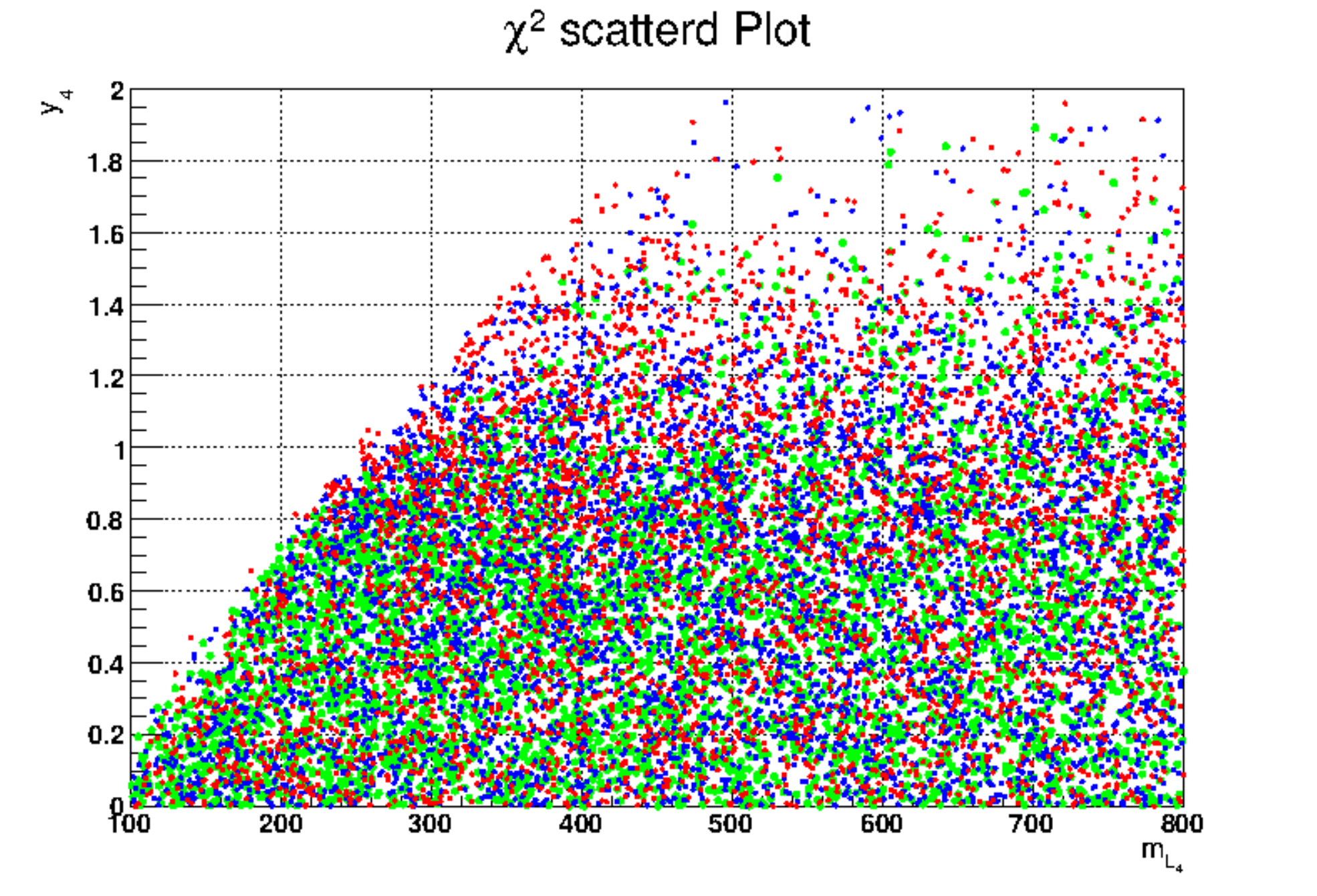}
\end{tabular}
\vskip 1mm
\caption{ Scatter plot of $\chi^2_{STU}$ over $m_{L_4}$ vs $y_4$ plane for vector like leptons with single doublet (left Fig.) and within two Higgs doublet (right Fig.) case.
The light green, blue and red dots correspond to $68$\%, $95$\% and $99$\% Confidence Limit (CL).
}
\label{fig.chisq}
\end{figure}

To study the effects of additional scalar  on Yukawa mixing and mass splittings between vector like species
we present the scatter plot of allowed region ($\chi^2 < 11.3$) over  $\Delta M_4
(=(m_{L_4} - m_{E_4})/(m_{L_4} + m_{E_4}))$ $vs.$ $y_4$ plane in Fig.~\ref{fig.splitting}  for pure vector like (left Fig.) and
with additional doublet (right Fig.) case.  The viable region prefers smaller Yukawa mixing ($y_4$)
and $m_{L_4}\approx m_{E_4}$ (maximum around 5\%) among SU(2) states in vector like case. Here one
can also draw an upper bound of $y_4 < 1.6$ in pure vector like case. However, all these conclusions of pure vector like case get substantially
modified under
presence of new scalar doublet. Now the parameter space allows almost all values of Yukawa mixing and  much greater
splittings (maximum around 15\%) among vector like states.
Finally in Fig.~\ref{fig.chisq} we present the  results corresponding to
$\chi^2_{STU}$ over $m_{L_4}-y_4$ plane. The much tighter constraints of  68\% CL can be satisfied all over $m_{L_4}-y_4$ plane in additional doublet case while they
are restricted only to some particular ranges of model parameter values in pure vector like case. However density of viable points do decreases in additional Higgs
doublet case as the contributions can also generate addditive effect which will offset them from permissible ranges.

\begin{figure}[!t]\centering
\begin{tabular}{c c}
\includegraphics[angle=0,width=75mm]{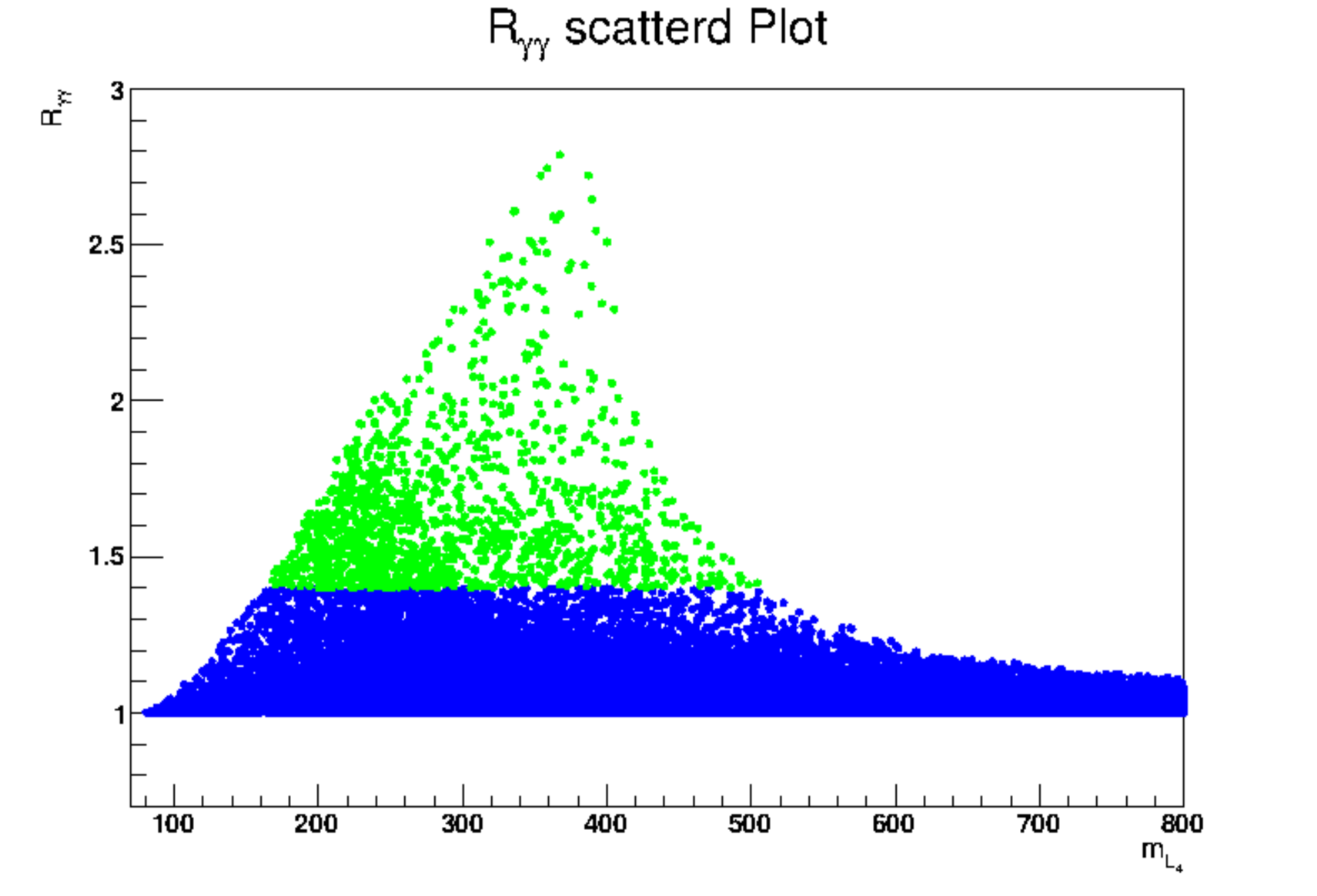} &
\includegraphics[angle=0,width=75mm]{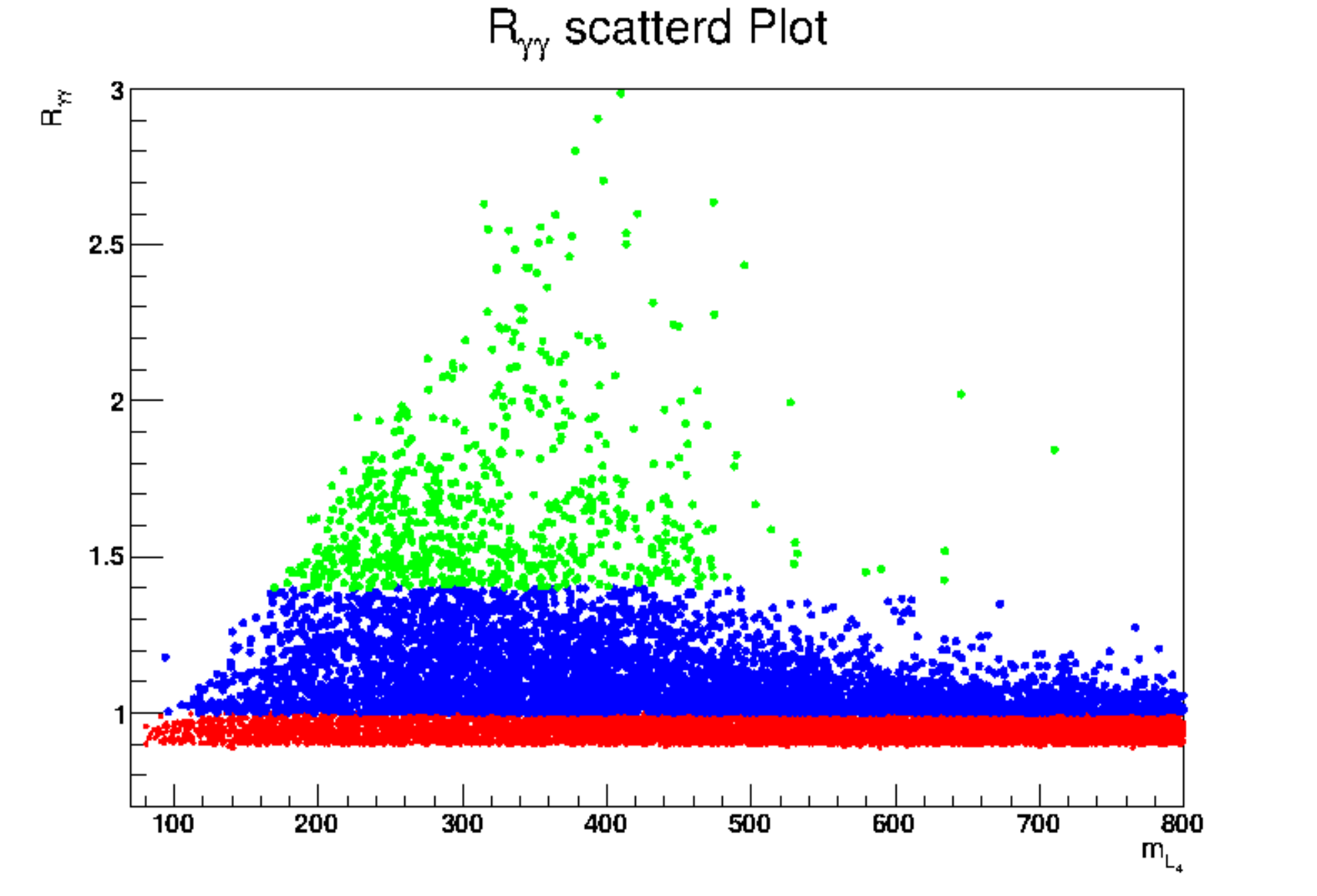}
\end{tabular}
\vskip 1mm
\caption{$R_{\gamma\gamma}$ vs $m_{L_4}$ plot  for vector like leptons with single doublet (left Fig.) and within two Higgs doublet (right Fig.) case.
Here red points correspond to $ R_{\gamma\gamma} < 1.0$, blue to $ 1.0 < R_{\gamma\gamma} < 1.4 $ and light green to $R_{\gamma\gamma} > 1.4$.
}
\label{fig1.rggtot}
\end{figure}

\begin{figure}[!t]\centering
\begin{tabular}{c c}
\includegraphics[angle=0,width=75mm]{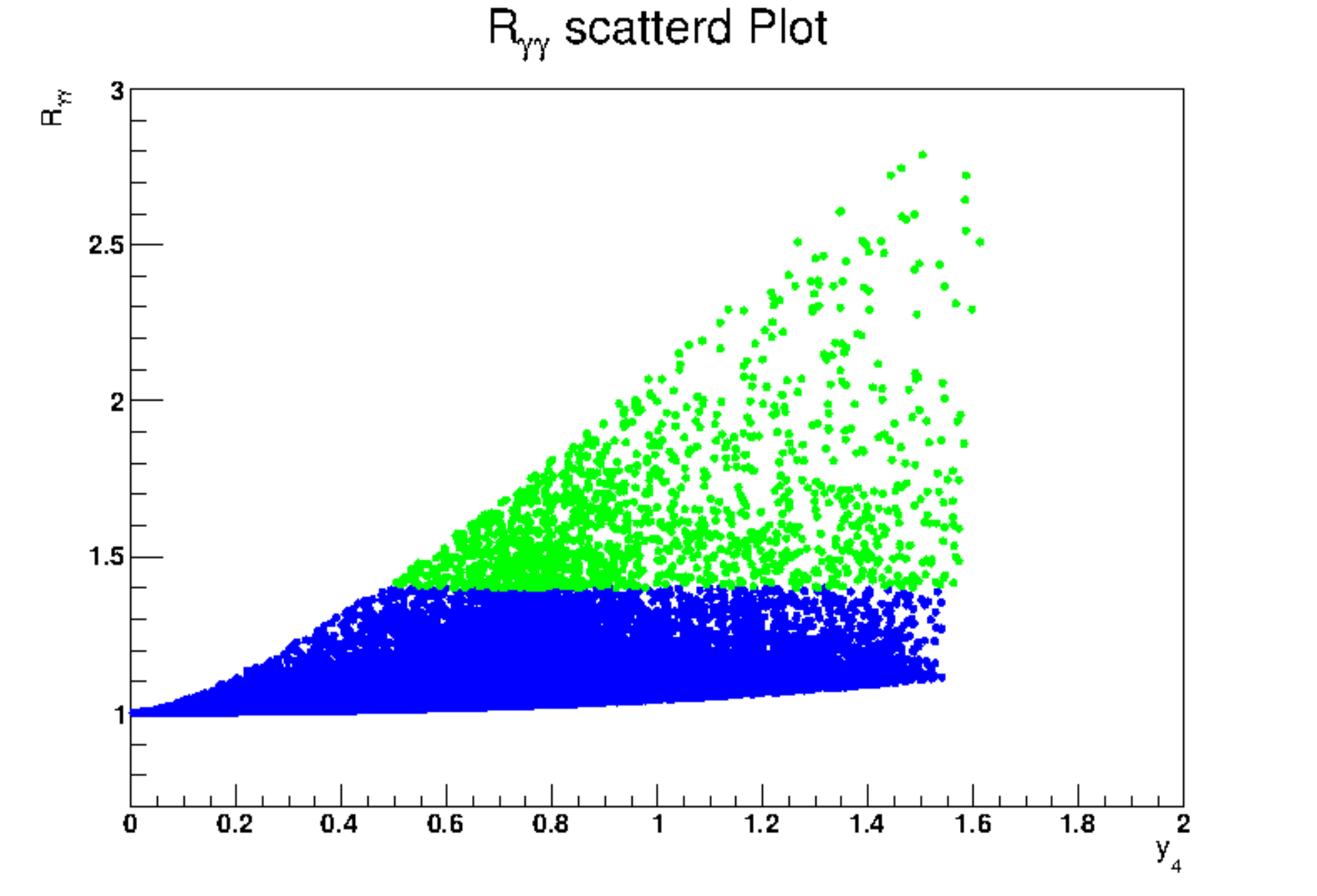} &
\includegraphics[angle=0,width=75mm]{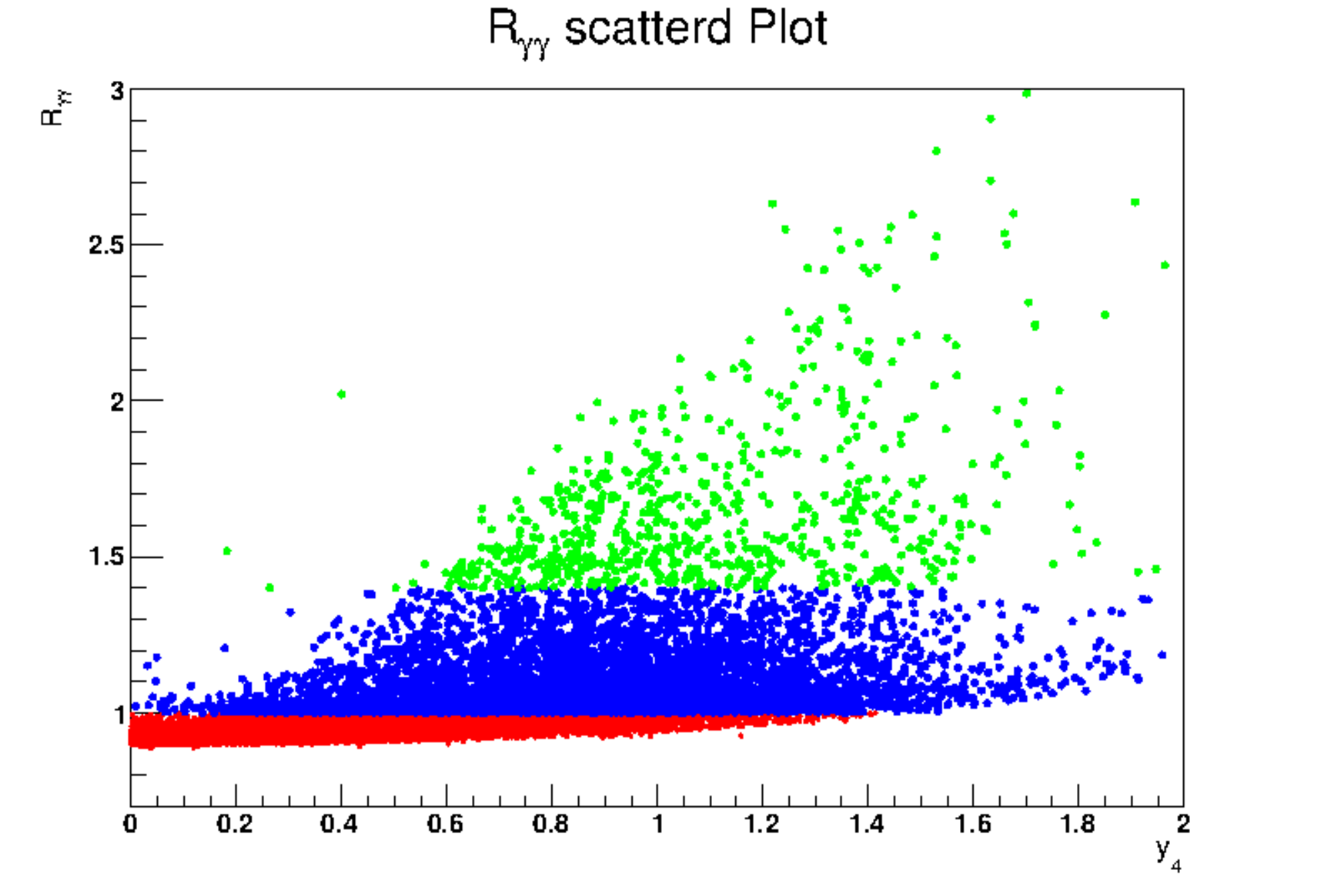}
\end{tabular}
\vskip 1mm
\caption{ $R_{\gamma\gamma}$ vs $y_4$ plot  for vector like leptons with single doublet (left Fig.) and within two Higgs doublet (right Fig.) case.
Here red points correspond to $ R_{\gamma\gamma} < 1.0$, blue to $ 1.0 < R_{\gamma\gamma} < 1.4 $ and light green to $R_{\gamma\gamma} > 1.4$.
}
\label{fig2.rggtot}
\end{figure}

\begin{figure}[!t]\centering
\begin{tabular}{c c}
\includegraphics[angle=0,width=75mm]{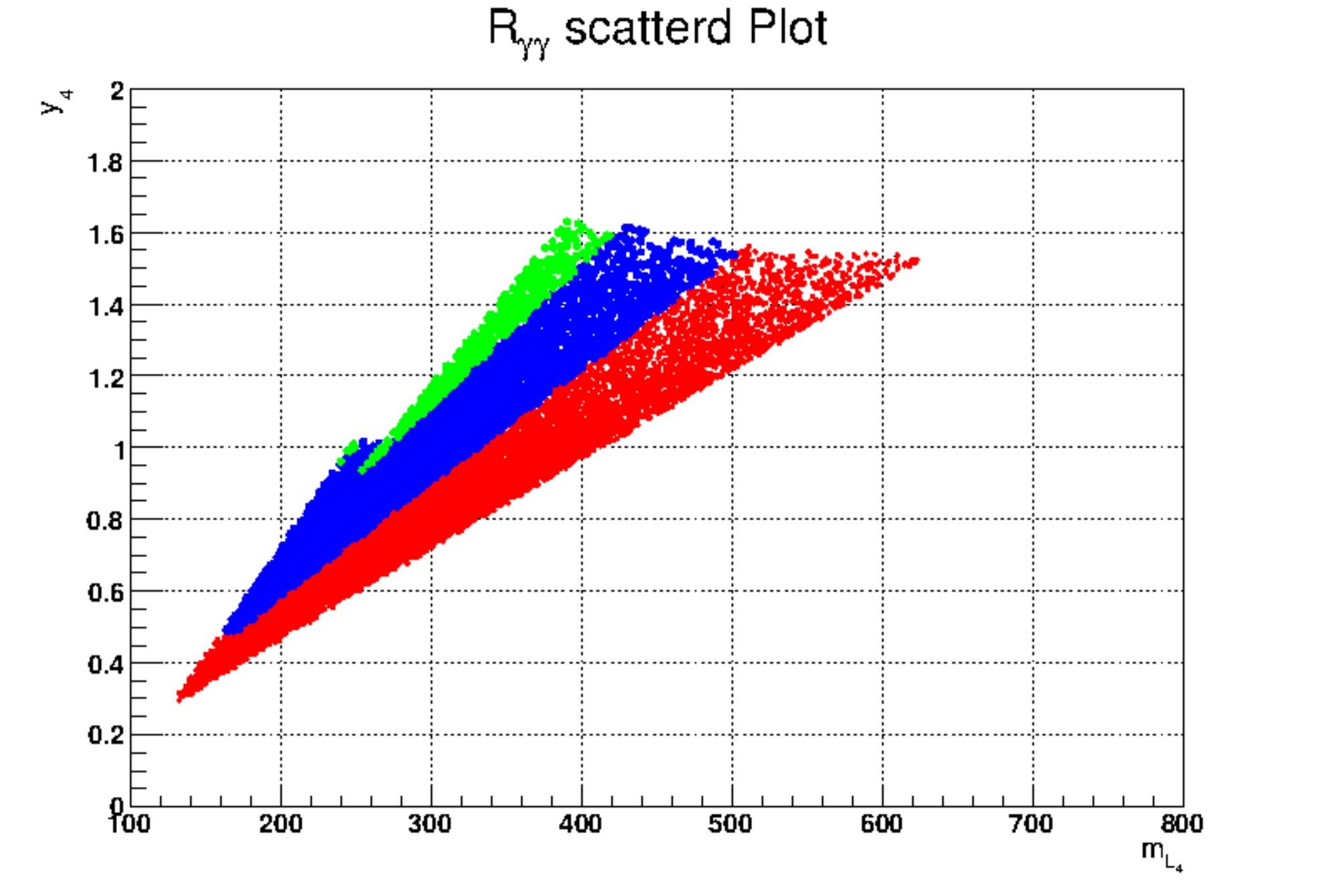} &
\includegraphics[angle=0,width=75mm]{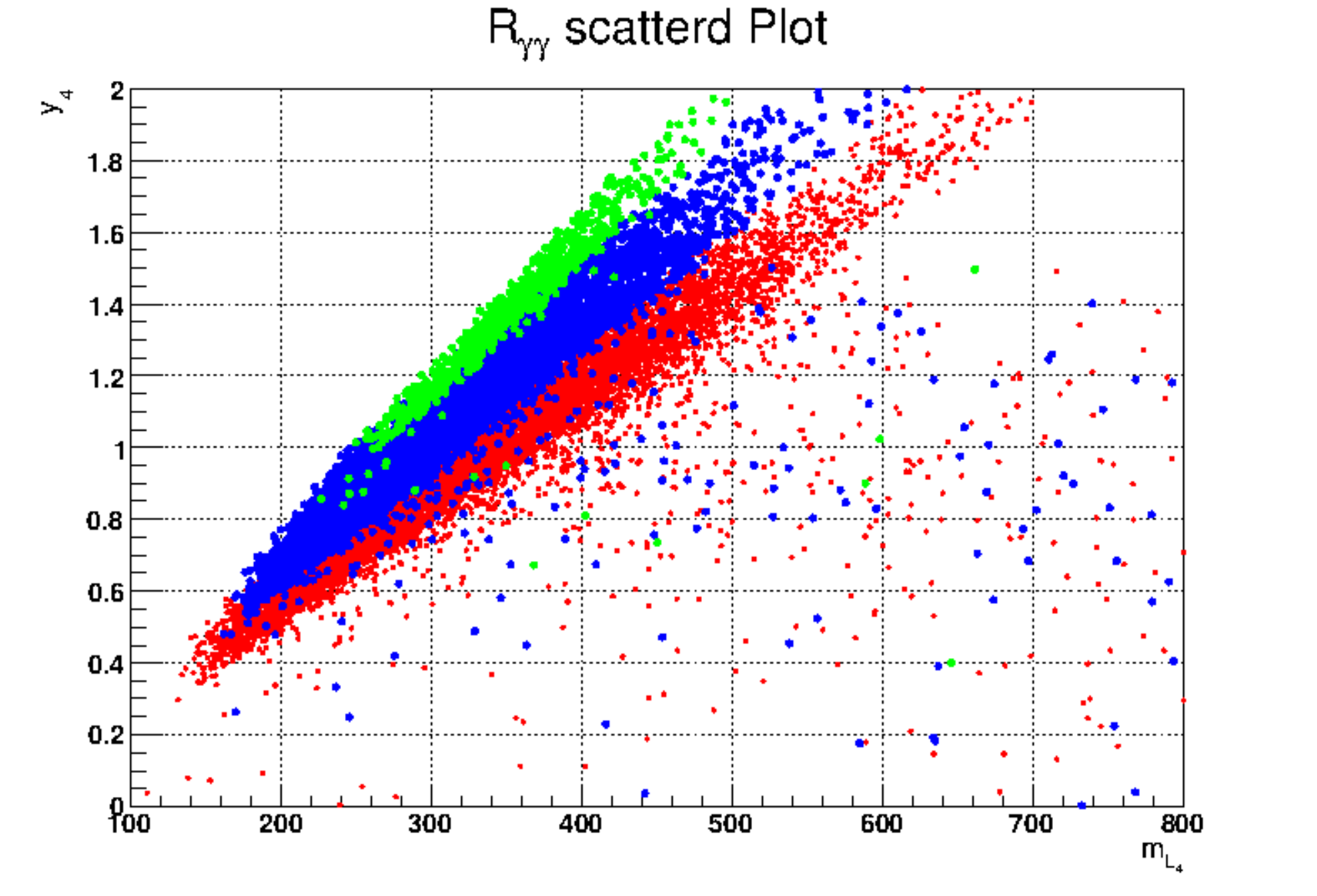}
\end{tabular}
\vskip 1mm
\caption{ Scattered plot of $R_{\gamma\gamma}$ over $m_{L_4}-y_4$ plane  for vector like leptons with single doublet (left Fig.) and within two Higgs doublet
(right Fig.) case. Here
red points correspond to $1.2 < R_{\gamma\gamma} < 1.4$, blue to $ 1.4 < R_{\gamma\gamma} < 2.0 $ and light green to $R_{\gamma\gamma} > 2.0$.
}
\label{fig.rggtot}
\end{figure}

Now we will investigate the situation with Higgs to diphoton rate in allowed parameter space. Here we imposed the condition $\chi^2_{STU} < 11.3$ on
parameter space which pertains to the 99\% CL on S, T and U parameters.
In Fig.~\ref{fig1.rggtot} we present the
$R_{\gamma\gamma}$ plot over $m_{L_4}$ parameter for both cases. In pure vector like case, $R_{\gamma\gamma} > 1.4$ can be obtained only in the range around
$m_{L_4} \in$ [150, 500]\,GeV, while for its corresponding
case with additional doublet similar enhancement is possible for
even higher  values of $m_{L_4}$ parameter. The another noticeable difference here is that in the case with additional doublet
it is also possible to generate cancellations between charged Higgs and fermionic contributions. Thus one can also get $R_{\gamma\gamma} < 1$ in some
region of parameter space which will support  CMS result. However, now parameter space will be much tighter compared to enhanced case.
The similar plots for $y_4$ are given in Fig.~\ref{fig2.rggtot}. As expected large enhancement prefers larger
values of Yukawa mixing while EWPD especially T parameter prefers it to be small. However, with two Higgs doublet case enhancement is  possible even for
all values of mixing values since now EWPD are easily satisfied here. Finally in  Fig.~\ref{fig.rggtot}
we presented $R_{\gamma\gamma}$ enhanced regions for two cases. Here red points correspond to lower enhancements ($1.4 > R_{\gamma\gamma} > 1.2$), blue correspond to
$2.0 > R_{\gamma\gamma} > 1.4$ and light green are for $R_{\gamma\gamma} > 2$. In single doublet case enhanced $R_{\gamma\gamma}$ region is
confined to narrow regions of model parameter values while with additional doublet the enhancement can be achieved in much larger parameter space. Thus the role of additional
doublet is two fold here. It
brings almost all the parameter space under 68\% CL of electroweak precision parameters and secondly it can also generate cancellations between two contributions in Higgs to 
gamma gamma channel and thus providing the possiblity to supress the decay rate.

\begin{figure}[!t]\centering
\begin{tabular}{c c}
\includegraphics[angle=0,width=75mm]{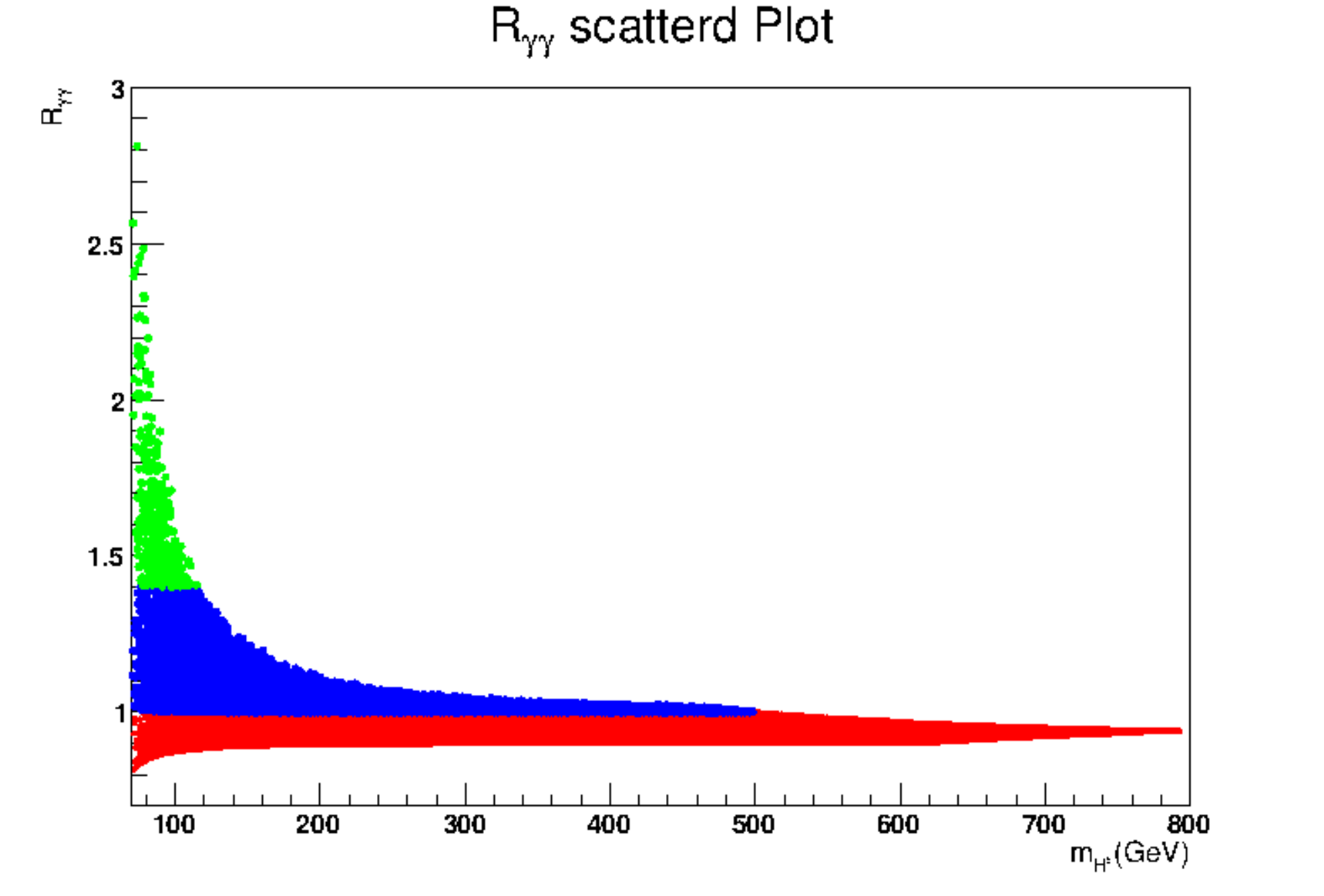} &
\includegraphics[angle=0,width=75mm]{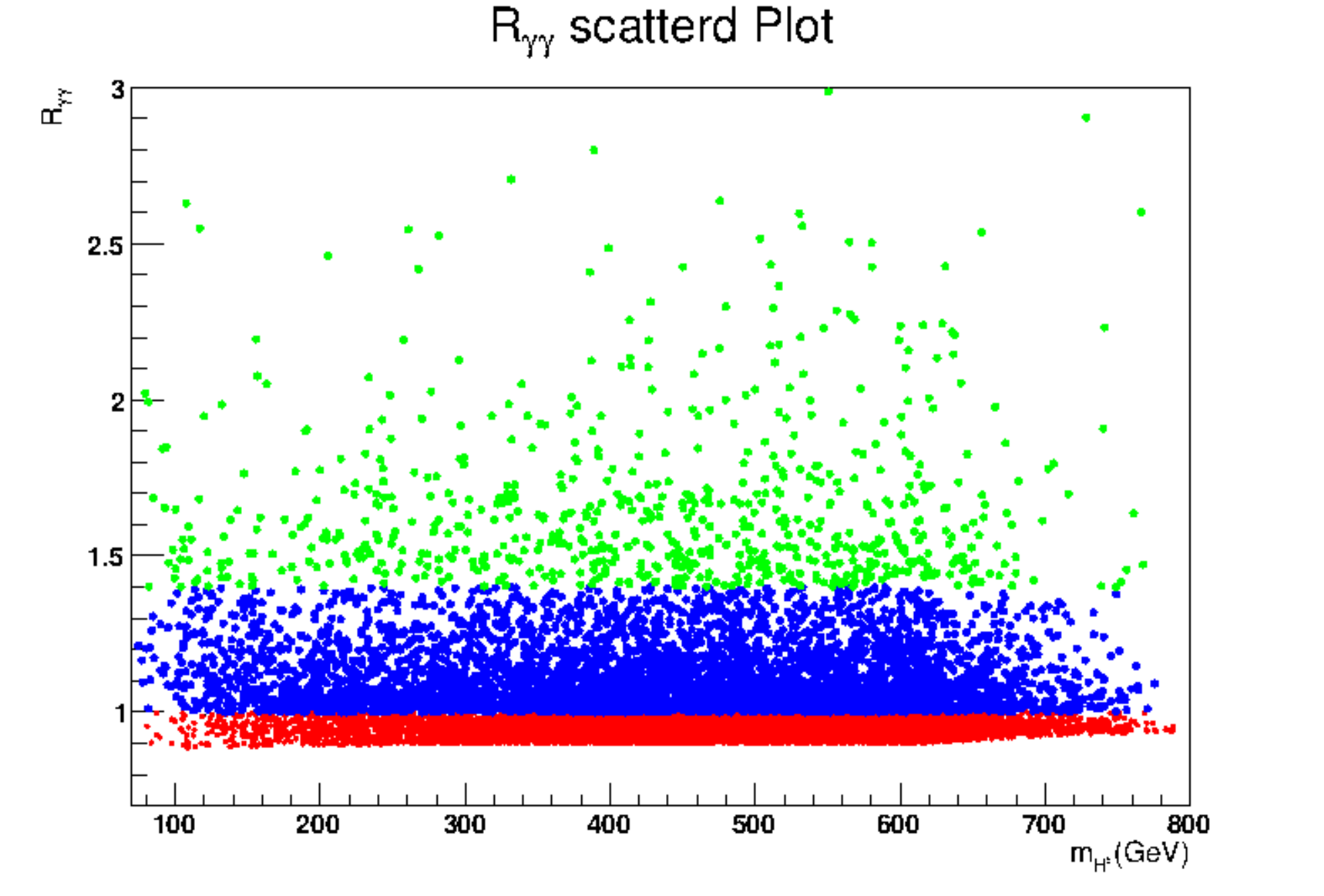}
\end{tabular}
\vskip 1mm
\caption{ $R_{\gamma\gamma}$ vs $m_{H^\pm}$ plot in 2HDM without (left Fig.) and with (right Fig.) vector like leptons.
Here red points correspond to $ R_{\gamma\gamma} < 1.0$, blue to $ 1.0 < R_{\gamma\gamma} < 1.4 $ and light green to $R_{\gamma\gamma} > 1.4$.
}
\label{fig4.rggtot}
\end{figure}

\begin{figure}[!t]\centering
\begin{tabular}{c c}
\includegraphics[angle=0,width=75mm]{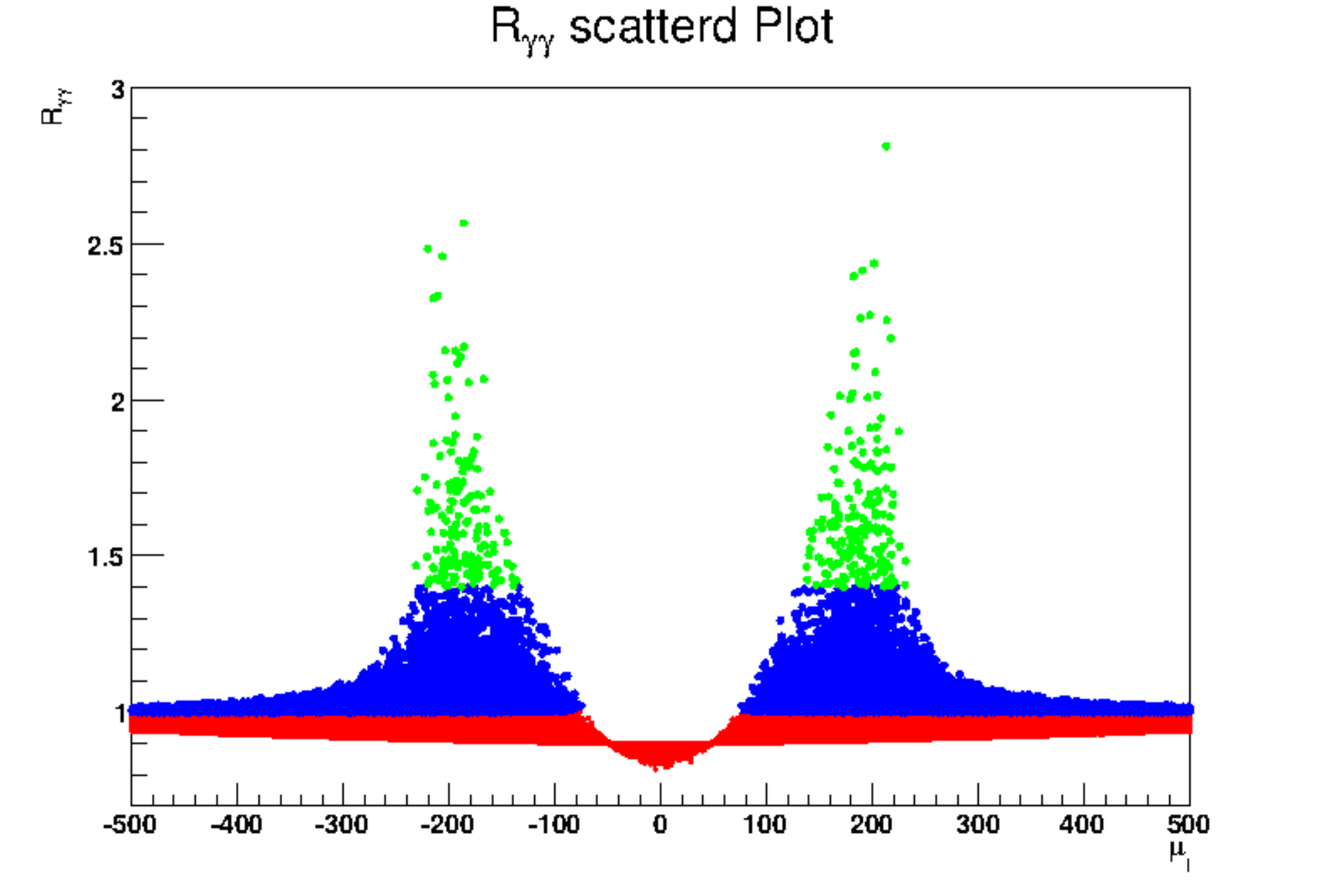} &
\includegraphics[angle=0,width=75mm]{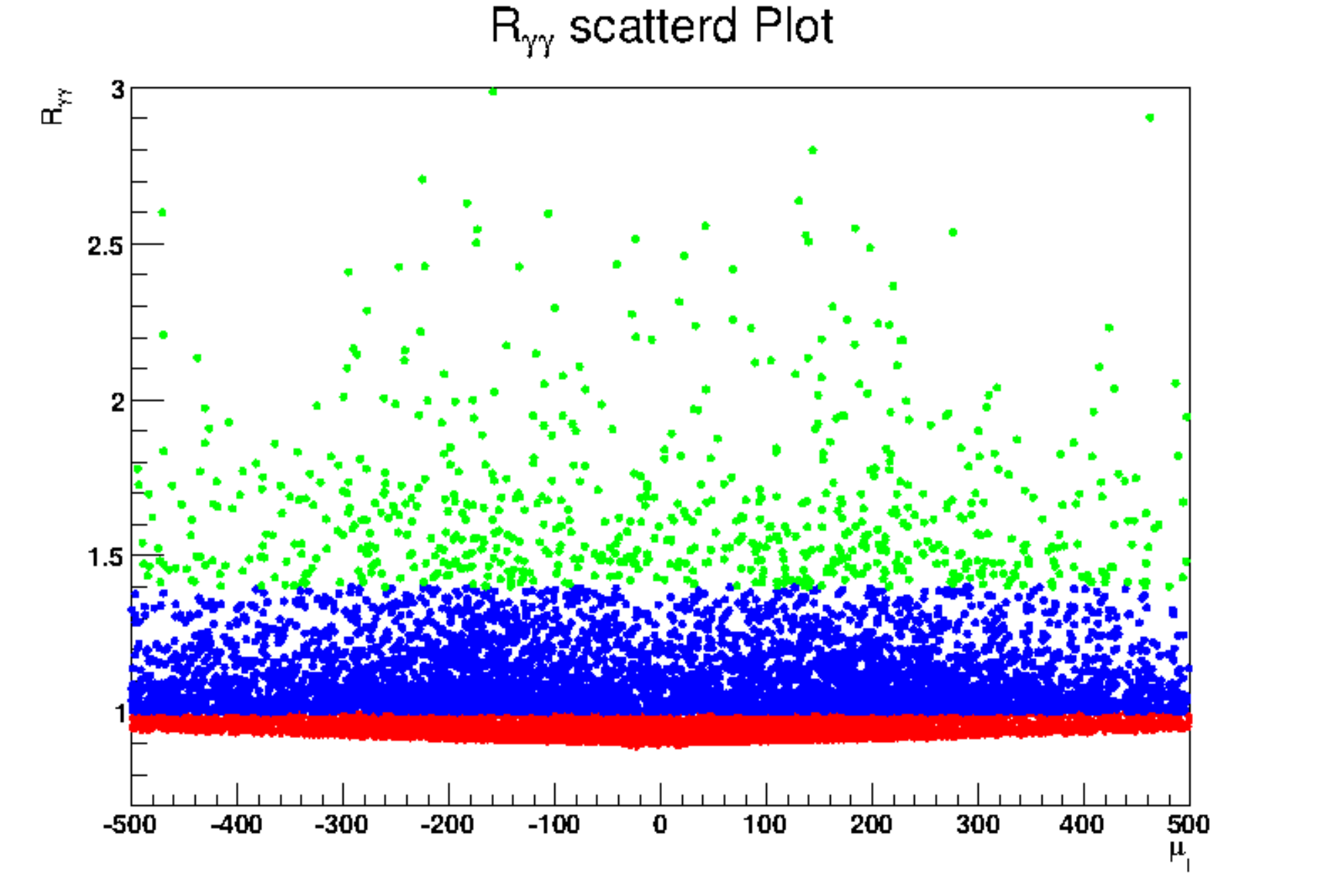}
\end{tabular}
\vskip 1mm
\caption{$R_{\gamma\gamma}$ vs $\mu_I$ plot in 2HDM without (left Fig.) and with (right Fig.) vector like leptons.
Here red points correspond to $ R_{\gamma\gamma} < 1.0$, blue to $ 1.0 < R_{\gamma\gamma} < 1.4 $ and light green to $R_{\gamma\gamma} > 1.4$.
}
\label{fig5.rggtot}
\end{figure}

As discussed in many recent studies \cite{hgmgmidm2,hgmgmidm3,hgmgmidm4} for Type I 2HDM  it is difficult to get larger enhancement
consistent with
ATLAS result due to stronger theoretical constraints. But vector like leptons
can impart a significant contribution on 2HDM parameter space. Thus now we will briefly comment on the effects of vector like leptons
in 2HDM parameter space pertaining  to $R_{\gamma\gamma}$ enhancement. In Fig.~\ref{fig4.rggtot}
we presented the $R_{\gamma\gamma}$ vs $m_{H^\pm}$ plot for 2HDM case and with vector like leptons case. In 2HDM case it's possible to have large
enhancement
only for lighter charged Higgs mass ($\approx$ 100\,GeV). Indeed one can draw an upper bound on charged Higgs mass i.e. $m_{H^\pm}< 125$\,GeV for
$R_{\gamma\gamma} > 1.4$. However, with vector like leptons this enhancement can be stretched to all values
of charged Higgs mass. In Fig.~\ref{fig5.rggtot} we presented the $R_{\gamma\gamma}$ vs $\mu_I$ for both cases. The plots
are symmetric under change of sign of $\mu_I$ as it the square of this parameter which enters into $hH^+H^-$ coupling. Moreover larger enhancement
can be achieved for all values of $\mu_I$ in 2HDM with vector like leptons unlike pure 2HDM case where it is confined to very narrow ranges
in $\mu_I$.

\begin{figure}[!t]\centering
\begin{tabular}{c c}
\includegraphics[angle=0,width=75mm]{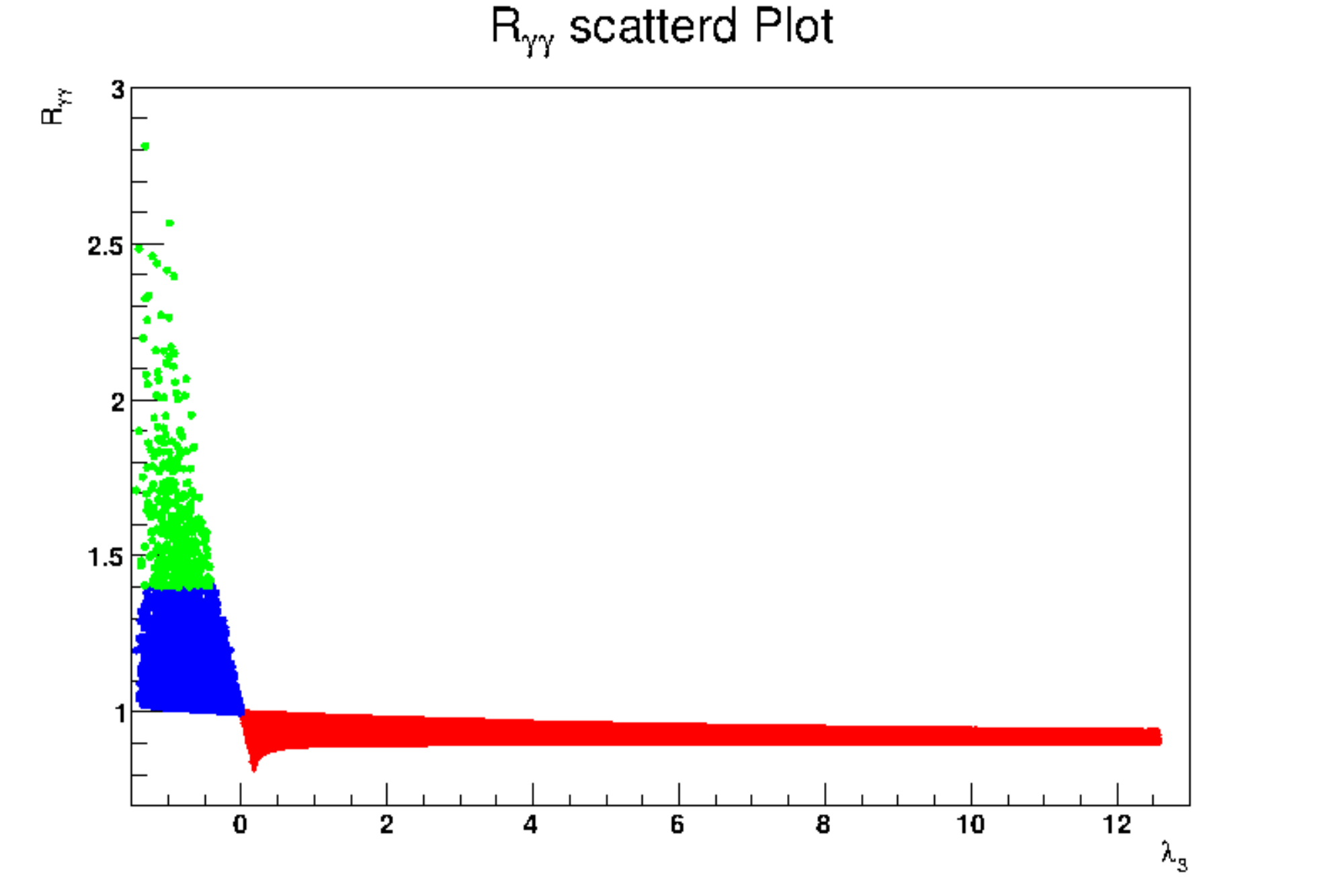} &
\includegraphics[angle=0,width=75mm]{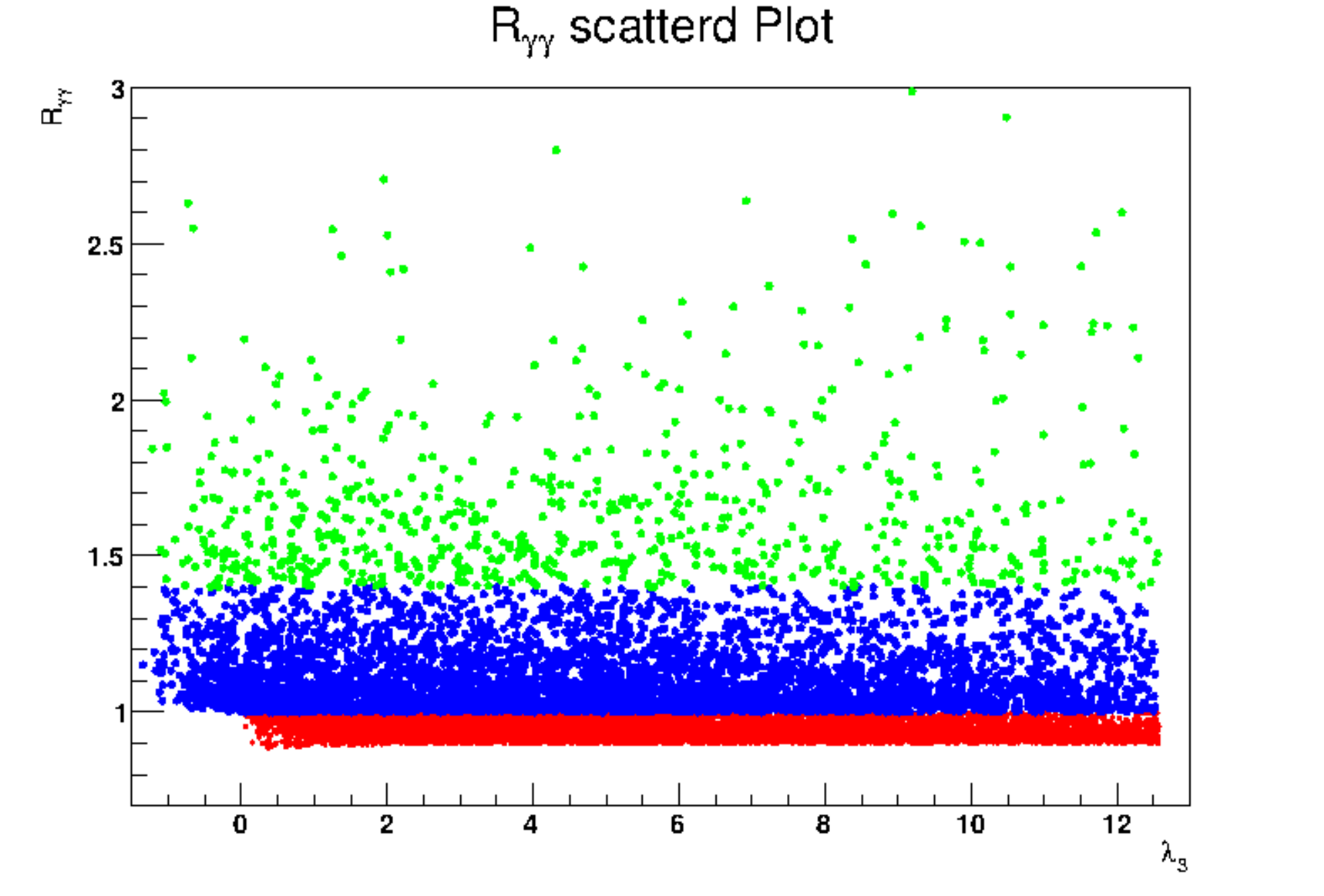}
\end{tabular}
\vskip 1mm
\caption{ $R_{\gamma\gamma}$ vs $\lambda_3$ plot in 2HDM without (left Fig.) and with (right Fig.) vector like leptons.
Here red points correspond to $ R_{\gamma\gamma} < 1.0$, blue to $ 1.0 < R_{\gamma\gamma} < 1.4 $ and light green to $R_{\gamma\gamma} > 1.4$.
}
\label{fig6.rggtot}
\end{figure}

\begin{figure}[!t]\centering
\begin{tabular}{c c}
\includegraphics[angle=0,width=75mm]{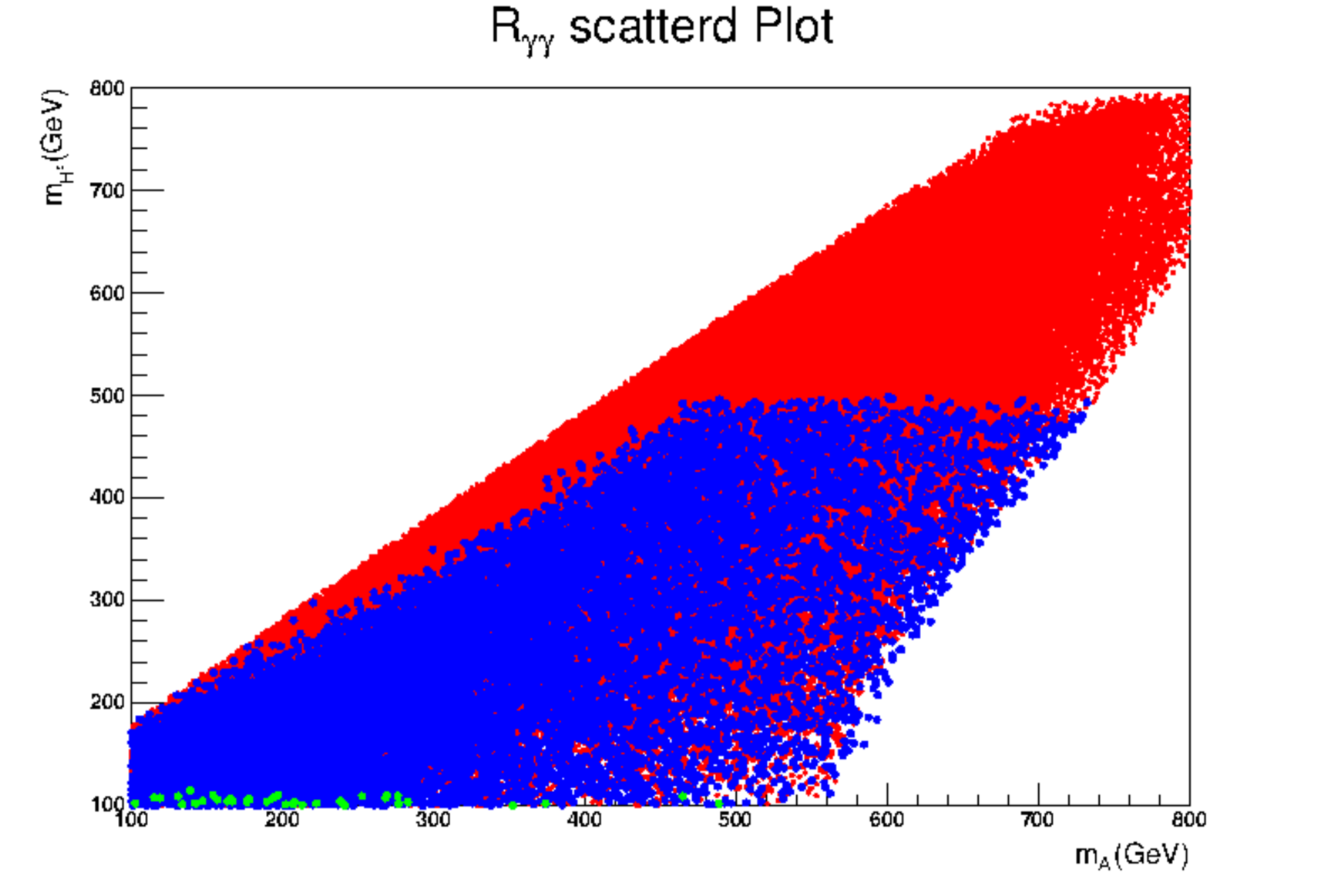} &
\includegraphics[angle=0,width=75mm]{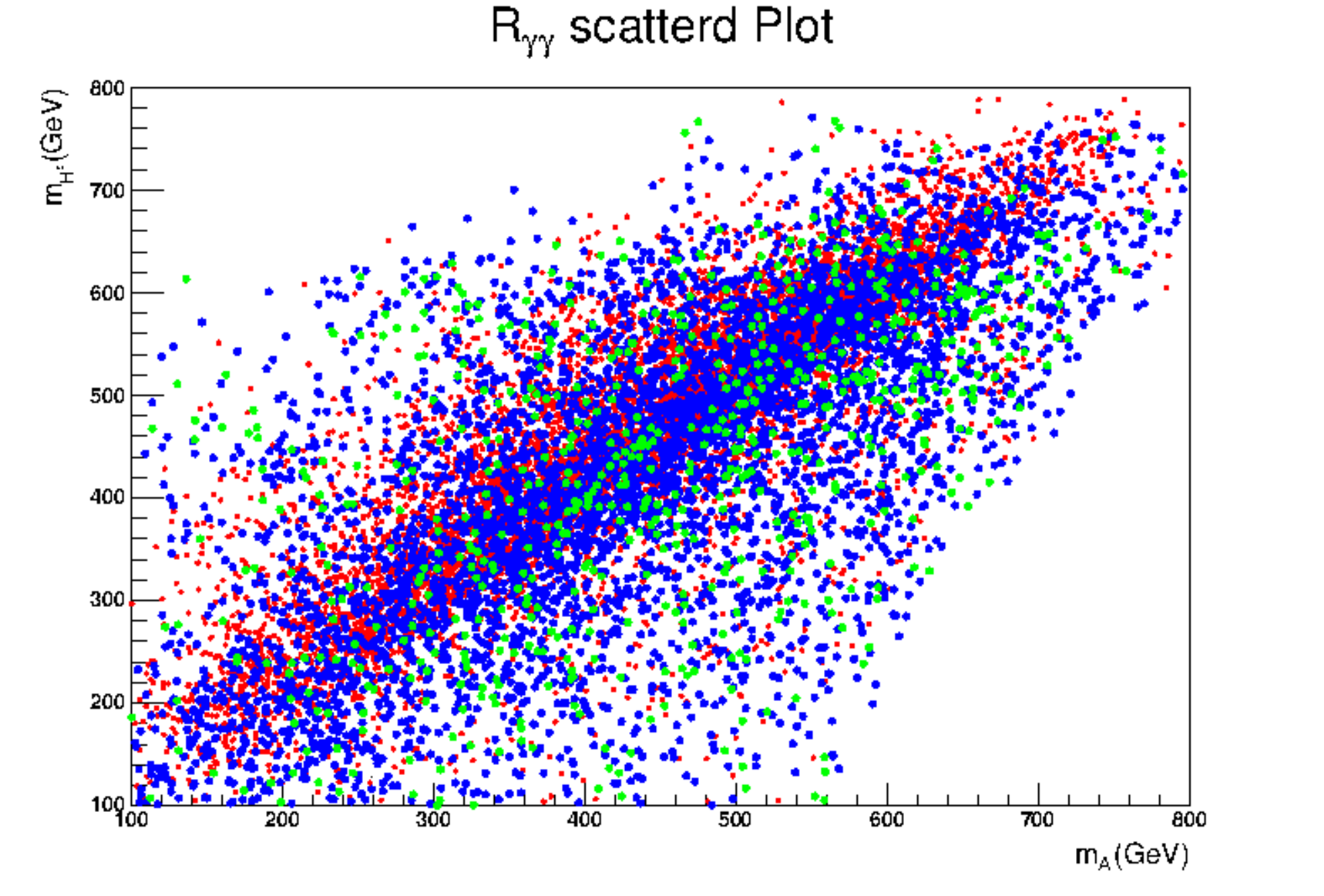}
\end{tabular}
\vskip 1mm
\caption{ $R_{\gamma\gamma}$ scattered plot over $m_{H^\pm}-m_A$ plane in 2HDM without (left Fig.) and with (right Fig.) vector like leptons. Here
red points correspond to $R_{\gamma\gamma} < 1$, blue to $ 1.0 < R_{\gamma\gamma} < 1.4 $ and light green to $R_{\gamma\gamma} > 1.4$.
}
\label{fig7.rggtot}
\end{figure}

In Fig.~\ref{fig6.rggtot} we presented the $R_{\gamma\gamma}$ vs. $\lambda_3$ for both cases. In pure 2HDM case enhancement
can be obtained only for $\lambda_3 < 0$ which corresponds to the constructive interference between W boson and charged Higgs($H^\pm$) contributions.
However with vector like leptons enhancement is even permitted for $\lambda_3 > 0$ since  contribution of vector like leptons becomes dominant
in decay rate. Finally in Fig.~\ref{fig7.rggtot} we presented the $R_{\gamma\gamma}$ over $m_{H^\pm}-m_A$ plane
for both cases. Here red points corresponds to $R_{\gamma} < 1$, blue points $R_{\gamma} \in [1,1.4]$ while light green to $R_{\gamma} > 1.4$. As elaborated earlier
it is difficult to get larger enhancement in 2HDM unless charged Higgs becomes very light while with vector like leptons it can be achieved all over this plane.

\section{Results and Discussion}

In this work motivated by the rich phenomenology of additional fermions
in extended Higgs sector we studied the influence of additional doublet on the SM supplemented with vector like
leptons. The similar studies for supersymmetric case are extensively
discussed \cite{susyvecferm} in literature. In particular we investigated  the case of the SM with one complete generation of
vector like leptons in context of inert doublet model \cite{idmall}.
The two doublets do not mix with each other and the lightest CP even state plays the role of
the SM Higgs. This will also be in accordance with the latest LHC results which show consistency with the SM Higgs boson.

Here first we studied the effects of new fermionic and scalar states on electroweak precision observables which are defined in terms of Peskin and
Takeuchi \cite{PeskinSTU} parameters (S, T, U). We scanned the parameter space of this model by imposing various constraints like vacuum stability,
perturbativity, unitarity along with these precision parameters. We showed with additional doublet it is possible to have cancellations
between scalar and fermionic contributions and
thus alleviating these constraints. This will in turn permit larger Yukawa mixing and mass splittings among vector like states.

Among  various properties of newly discovered scalar resonance, its loop induced decays serves as a much sensitive probe of new physics. Here BSM fields
that couple to Higgs can challenge the expectations of the SM by circulating in its loop decay diagrams. In this study we focused on Higgs to gamma gamma channel as
signature of BSM physics.  ATLAS reported an excess in this channel with signal strength $\sigma/\sigma_{SM} = 1.65 \pm 0.24(stat)^{+0.25}_{-0.18}(syst)$ while CMS number
comes down from its previous results  $\sigma/\sigma_{SM}= 1.56 \pm 0.43$ to
$\sigma/\sigma_{SM}= 1.11 \pm 0.31$ with cut based events and $0.78\pm 0.27$ with selected and categorized events. Thus results in this
channel doesn't seem to be entirely consistent with the SM and thus provide ample space for new physics.

In this scenario, the charged fermionic and scalar fields contribute in the loop induced decays of Higgs and thus can give  signatures
which are different from single Higgs doublet case. We identified various regions corresponding to Higgs to gamma gamma decay in allowed parameter space. We also
compared and  contrasted them
with vector like leptons in single Higgs doublet and  two Higgs doublet model (2HDM) case.
As discussed in numerical section the role of additional doublet is two fold here. It
brings almost all the parameter space under 68\% CL of electroweak precision parameters unlike single doublet case where only a very constrained region comes under
these stricter constraints. Moreover it can also generate cancellations between new fermionic and scalar contribution of Higgs to gamma gamma channel.
This will in turn even help in suppressing $R_{\gamma\gamma}$ as indicated by CMS.  However, the enhancement regions consistent with ATLAS in this model are now stretched to much larger
range of model parameters. Moreover in vice versa case vector like leptons can have significant effect on 2HDM where
excess in $R_{\gamma\gamma}$ consistent with ATLAS can only obtained for very narrow ranges of model parameters. Now the enhancement can be obtained for almost all values of 2HDM parameters.
A more detailed analysis of all channels in this model will be presented somewhere else. Thus in conclusion additional doublet impart significant effect on the
parameter space of the SM with vector like leptons. These scenarios not only provide rich phenomenology but also have implications quite different from simple cases.

\bigskip

\noindent {\bf{Acknowledgements}} \\
We thank Ardy Mustafa for collaborating in the initial stages of this work. The work of SKG and CSK is supported by the National Research Foundation
of Korea (NRF) grant funded by Korea government of the Ministry of Education, Science and Technology (MEST) (Grant No. 2011-0017430 and Grant No. 2011-0020333).

\bigskip

\begin{appendix}

\section{Loop Functions}\label{loopfuncts}
The various 1-loop functions which  appear in the  calculation of decay width
$\Gamma(h \to \gamma\gamma)$ are given as:
\begin{eqnarray}
&& V_1(x) = 2 + 3 x + 3x(2-x)f(x),\nonumber\\
&& F_{1/2}(x)= -2 x[1+(1-x)f(x)],\nonumber\\
&& C_0(x) = x[1-xf(x)]
\end{eqnarray}
with
\begin{equation}
f(x) = \left\{ \begin{array}{lr}
[\sin^{-1}(1/\sqrt{x})]^2, & x \geq 1 \\
-\frac{1}{4} [\ln(\eta_+/\eta_-) - i \pi]^2, & \, x < 1
\end{array}  \right.
\end{equation}
and
\begin{equation}
x_i = 4 m_i^2 / m_h^2, \qquad \eta_{\pm} = 1 \pm \sqrt{1-x}.
\end{equation}

\section{S, T and U parameters}

Here we are giving the contribution of scalar and fermionic sector
which is already available in literature.

\subsection{Doublet Contribution}
The one loop contribution to the oblique parameters(S and T) in  inert doublet model
is given by\cite{STIDM,STUcurrent}:
\begin{eqnarray}
 T &=& \frac{1}{16\pi^2 \alpha v^2}\Bigg[ F(m^2_{H^\pm}, m^2_{A})
+ F(m^2_{H^\pm}, m^2_{S}) - F(m^2_{A}, m^2_{S})\Bigg],
\end{eqnarray}
and
\begin{eqnarray}
 S = \frac{1}{2\pi}\Bigg[ \frac{1}{6}\log(\frac{m^2_{S}}{m^2_{H^\pm}}) -
  \frac{5}{36} + \frac{m^2_{S} m^2_{A}}{3(m^2_{A}-m^2_{S})^2} +
\frac{m^4_A (m^2_{A}-3m^2_{S})}{6(m^2_{A}-m^2_{S})^3} \log(\frac{m^2_{A}}{m^2_{S}})\Bigg],
\end{eqnarray}
where the function $F$ is defined by
\begin{equation}
	F(x,y) = \left\{ \begin{array}{lr}
		\frac{x+y}{2} - \frac{xy}{x-y}\log(\frac{x}{y}), & x\neq y. \\
		0, & \, x=y.
		\end{array}  \right.
\end{equation}

\subsection{Vector like Fermion Contribution}

The contributions to the gauge boson two point functions from fermion loops parametrized by the interaction
\begin{equation}
{{\cal L}=\overline f}_1
\biggl(g_{LX}^{f_1f_2} P_L
+g_{RX}^{f_1f_2} P_R\biggr)\gamma_\mu f_2 V^\mu\,,
\end{equation}
for $V=W,Z,\gamma$ is given by~\cite{VecEwpr}
\begin{eqnarray}
\Pi_{XY}(p^2,m_1^2, m_2^2)&=&
-{N_c\over 16 \pi^2}\biggl\{
{2\over 3}\biggl(g_{LX}^{f_1f_2}g_{LY}^{f_1f_2}+
g_{RX}^{f_1f_2}g_{RY}^{f_1f_2}\biggr)
\biggl[m_1^2+m_2^2-{p^2\over 3}-
\biggl(A_0(m_1^2)+A_0(m_2^2)\biggr)
\nonumber \\
&&+{m_1^2-m_2^2\over 2 p^2}
\biggl(A_0(m_1^2)-A_0(m_2^2)\biggr)
+{2p^4-p^2(m_1^2+m_2^2)-(m_1^2-m_2^2)^2\over 2 p^2}
B_0(p^2, m_1^2,m_2^2)\biggr]
\nonumber \\
&&
+2m_1m_2\biggl(g_{LX}^{f_1f_2}g_{RY}^{f_1f_2}+
g_{RX}^{f_1f_2}g_{LY}^{f_1f_2}\biggr)B_0(p^2,m_1^2,m_2^2)\biggr\}.
\end{eqnarray}

Here $N_c$ is the number of color degrees of freedom. In the limit of zero external momentum two point function goes to
\begin{eqnarray}
\Pi_{XY}(0,m_1^2, m_2^2)&=&
-{N_c\over 16 \pi^2}\biggl\{
{2\over 3}\biggl(g_{LX}^{f_1f_2}g_{LY}^{f_1f_2}+
g_{RX}^{f_1f_2}g_{RY}^{f_1f_2}\biggr)
\biggl[m_1^2+m_2^2 -
\biggl(A_0(m_1^2)+A_0(m_2^2)\biggr)
\nonumber \\
&&-{m_1^2+ m_2^2\over 2 }
B_0(0,m_1^2, m_2^2)
-{(m_1^2-m_2^2)^2\over 2 }
B_0^{'}(0, m_1^2,m_2^2)\biggr]
\nonumber \\
&&
+2m_1m_2\biggl(g_{LX}^{f_1f_2}g_{RY}^{f_1f_2}+
g_{RX}^{f_1f_2}g_{LY}^{f_1f_2}\biggr)B_0(0,m_1^2,m_2^2)\biggr\},
\end{eqnarray}
where
\begin{eqnarray}
A_0(m^2)&=&\biggl({4\pi\mu^2\over m^2}\biggr)^\epsilon
\Gamma(1+\epsilon)
\biggl({1\over \epsilon}+1\biggr) m^2 ,
\nonumber \\
B_0(p^2,m_1^2,m_2^2)&=&\biggl({4\pi\mu^2
\over m_2^2}\biggr)^\epsilon\Gamma(1+\epsilon)
\left[{1\over \epsilon} -f_1(p^2,m_1^2,m_2^2)\right] ,\nonumber\\
B_0^{'}(p^2,m_1^2,m_2^2)&=& \frac{\partial}{\partial p^2} B_0(p^2,m_1^2,m_2^2) ,
\end{eqnarray}
and
\begin{equation}
f_1(p^2, m_1^2,m_2^2)=\int_0^1 dx \log\biggl(x+{m_1^2(1-x)-p^2x(1-x)\over m_2^2}
\biggr)\, .
\end{equation}

Using above expressions one can easily calculate S, T and U parameters from
Eq.\,(\ref{eq:STU}) for vector like fermions.


\end{appendix}


\begin{thebibliography}{10}

\bibitem{ATLAS:2012ae}
{\bf ATLAS Collaboration} Collaboration, G.~Aad et~al., {\it {Combined search
  for the Standard Model Higgs boson using up to 4.9 fb-1 of pp collision data
  at sqrt(s) = 7 TeV with the ATLAS detector at the LHC}},  {\em Phys.Lett.}
  {\bf B710} (2012) 49--66, [\href{http://xxx.lanl.gov/abs/1202.1408}{{\tt
  arXiv:1202.1408}}].

\bibitem{Chatrchyan:2012tx}
{\bf CMS Collaboration} Collaboration, S.~Chatrchyan et~al., {\it {Combined
  results of searches for the standard model Higgs boson in pp collisions at
  sqrt(s) = 7 TeV}},  {\em Phys.Lett.} {\bf B710} (2012) 26--48,
  [\href{http://xxx.lanl.gov/abs/1202.1488}{{\tt arXiv:1202.1488}}].



\bibitem{MoroindHiggs}
The Atlas Collaboration, ATLAS-CONF-2013-029,\,
http://cds.cern.ch/record/1527124/files/ATLAS-CONF-2013-029.pdf;\\
The Atlas Collaboration, ATLAS-CONF-2013-013,\,
http://cds.cern.ch/record/1523699/files/ATLAS-CONF-2013-013.pdf;\\
The Atlas Collaboration, ATLAS-CONF-2013-031,\,
http://cds.cern.ch/record/1527127/files/ATLAS-CONF-2013-031.pdf;\\
The CMS Collaboration, HIG-13-002-pas,\,
http://cds.cern.ch/record/1523767/files/HIG-13-002-pas.pdf;\\
The CMS Collaboration, HIG-13-003-pas,\,
http://cds.cern.ch/record/1523673/files/HIG-13-003-pas.pdf


\bibitem{chiral4G}
  P.~H.~Frampton, P.~Q.~Hung and M.~Sher,
  Phys.\ Rept.\  {\bf 330}, 263 (2000)
  [hep-ph/9903387];
  B.~Holdom, W.~S.~Hou, T.~Hurth, M.~L.~Mangano, S.~Sultansoy and G.~Unel,
  PMC Phys.\ A {\bf 3}, 4 (2009)
  [arXiv:0904.4698 [hep-ph]];
   J.~Erler and P.~Langacker,
  Phys.\ Rev.\ Lett.\  {\bf 105}, 031801 (2010)
  [arXiv:1003.3211 [hep-ph]];
   C.~Anastasiou, R.~Boughezal and E.~Furlan,
  JHEP {\bf 1006}, 101 (2010)
  [arXiv:1003.4677 [hep-ph]];
  C.~Anastasiou, S.~Buehler, E.~Furlan, F.~Herzog and A.~Lazopoulos,
  Phys.\ Lett.\ B {\bf 702}, 224 (2011)
  [arXiv:1103.3645 [hep-ph]];
  O.~Eberhardt, G.~Herbert, H.~Lacker, A.~Lenz, A.~Menzel, U.~Nierste and M.~Wiebusch,
  Phys.\ Rev.\ Lett.\  {\bf 109}, 241802 (2012)
  [arXiv:1209.1101 [hep-ph]].
  
 
  
  




\bibitem{VecEwpr}
  M.~-C.~Chen and S.~Dawson,
  Phys.\ Rev.\ D {\bf 70}, 015003 (2004)
  [hep-ph/0311032];
  G.~Cynolter and E.~Lendvai,
  Eur.\ Phys.\ J.\ C {\bf 58}, 463 (2008)
  [arXiv:0804.4080 [hep-ph]];
  S.~Dawson and E.~Furlan,
  Phys.\ Rev.\ D {\bf 86}, 015021 (2012)
  [arXiv:1205.4733 [hep-ph]].





\bibitem{vecferms}
   G.~Cacciapaglia, A.~Deandrea, D.~Harada and Y.~Okada,
  JHEP {\bf 1011}, 159 (2010)
  [arXiv:1007.2933 [hep-ph]];
  G.~Cacciapaglia, A.~Deandrea, L.~Panizzi, N.~Gaur, D.~Harada and Y.~Okada,
  JHEP {\bf 1203}, 070 (2012)
  [arXiv:1108.6329 [hep-ph]];
  C.~Arina, R.~N.~Mohapatra and N.~Sahu,
  Phys.\ Lett.\ B {\bf 720}, 130 (2013)
  [arXiv:1211.0435 [hep-ph]];
  R.~Dermisek and A.~Raval,
  arXiv:1305.3522 [hep-ph].







\bibitem{4G2HDM}
  S.~Bar-Shalom, S.~Nandi and A.~Soni,
  Phys.\ Rev.\ D {\bf 84}, 053009 (2011)
  [arXiv:1105.6095 [hep-ph]];
 S.~Bar-Shalom, S.~Nandi and A.~Soni,
  Phys.\ Lett.\ B {\bf 709}, 207 (2012)
  [arXiv:1112.3661 [hep-ph]];
N.~Chen and H.~-J.~He,
  JHEP {\bf 1204}, 062 (2012)
  [arXiv:1202.3072 [hep-ph]];
  L.~Bellantoni, J.~Erler, J.~J.~Heckman and E.~Ramirez-Homs,
  Phys.\ Rev.\ D {\bf 86}, 034022 (2012)
  [arXiv:1205.5580 [hep-ph]];
 S.~Bar-Shalom, M.~Geller, S.~Nandi and A.~Soni,
  arXiv:1208.3195 [hep-ph];
M.~Geller, S.~Bar-Shalom, G.~Eilam and A.~Soni,
  Phys.\ Rev.\ D {\bf 86}, 115008 (2012)
  [arXiv:1209.4081 [hep-ph]].



 
 
\bibitem{vecferm1} 
  C.~Englert and M.~McCullough,
  arXiv:1303.1526 [hep-ph].

\bibitem{vecferm2}
  J.~Kearney, A.~Pierce and N.~Weiner,
  Phys.\ Rev.\ D {\bf 86}, 113005 (2012)
  [arXiv:1207.7062 [hep-ph]].

\bibitem{vecferm3}
  L.~G.~Almeida, E.~Bertuzzo, P.~A.~N.~Machado and R.~Z.~Funchal,
  JHEP {\bf 1211}, 085 (2012)
  [arXiv:1207.5254 [hep-ph]].

\bibitem{vecferm4}
  A.~Joglekar, P.~Schwaller and C.~E.~M.~Wagner,
  JHEP {\bf 1212}, 064 (2012)
  [arXiv:1207.4235 [hep-ph]].

\bibitem{vecferm5}
  N.~Arkani-Hamed, K.~Blum, R.~T.~D'Agnolo and J.~Fan,
  JHEP {\bf 1301}, 149 (2013)
  [arXiv:1207.4482 [hep-ph]].









\bibitem{susyvecferm}
  K.~S.~Babu, I.~Gogoladze, M.~U.~Rehman and Q.~Shafi,
  Phys.\ Rev.\ D {\bf 78}, 055017 (2008)
  [arXiv:0807.3055 [hep-ph]];
  S.~P.~Martin,
  Phys.\ Rev.\ D {\bf 81}, 035004 (2010)
  [arXiv:0910.2732 [hep-ph]];
  S.~P.~Martin,
  Phys.\ Rev.\ D {\bf 82}, 055019 (2010)
  [arXiv:1006.4186 [hep-ph]];
  S.~P.~Martin,
  Phys.\ Rev.\ D {\bf 83}, 035019 (2011)
  [arXiv:1012.2072 [hep-ph]].


\bibitem{susyvecfermRec}
 W.~-Z.~Feng and P.~Nath,
  arXiv:1303.0289 [hep-ph];
  A.~Joglekar, P.~Schwaller and C.~E.~M.~Wagner,
  arXiv:1303.2969 [hep-ph].



\bibitem{idmall}
  N.~G.~Deshpande and E.~Ma,
  Phys.\ Rev.\ D {\bf 18}, 2574 (1978);
  Q.~-H.~Cao, E.~Ma and G.~Rajasekaran,
  Phys.\ Rev.\ D {\bf 76}, 095011 (2007)
  [arXiv:0708.2939 [hep-ph]];
  A.~Goudelis, B.~Herrmann and O.~Stal,
  arXiv:1303.3010 [hep-ph].



\bibitem{PeskinSTU}
  M.~E.~Peskin and T.~Takeuchi,
  Phys.\ Rev.\ D {\bf 46}, 381 (1992).



\bibitem{ATLAShtogmgm}
  G.~Aad {\it et al.}  [ATLAS Collaboration],
  Phys.\ Lett.\ B {\bf 716}, 1 (2012)
  [arXiv:1207.7214 [hep-ex]].

\bibitem{CMShtogmgm}
  S.~Chatrchyan {\it et al.}  [CMS Collaboration],
  Phys.\ Lett.\ B {\bf 716}, 30 (2012)
  [arXiv:1207.7235 [hep-ex]].



\bibitem{ATLAShtogmgmnew}
The Atlas Collaboration, ATLAS-CONF-2013-012,\,
http://cds.cern.ch/record/1523698/files/ATLAS-CONF-2013-012.pdf

\bibitem{CMShtogmgmnew}
The CMS Collaboration, Hig13001TWiki,\,
https://twiki.cern.ch/twiki/bin/genpdf/CMSPublic/Hig13001TWiki


\bibitem{Gunionetal2HDM}
  J.~F.~Gunion and H.~E.~Haber,
  Phys.\ Rev.\ D {\bf 67}, 075019 (2003)
  [hep-ph/0207010].


\bibitem{hgmgmidm2}
  P.~Posch,
  Phys.\ Lett.\ B {\bf 696}, 447 (2011)
  [arXiv:1001.1759 [hep-ph]].


\bibitem{2HDMsymbrk}
  N.~G.~Deshpande and E.~Ma,
  Phys.\ Rev.\ D {\bf 18}, 2574 (1978);
  M.~Sher,
  Phys.\ Rept.\  {\bf 179}, 273 (1989).



\bibitem{unitarity1}
  B.~W.~Lee, C.~Quigg and H.~B.~Thacker,
  Phys.\ Rev.\ D {\bf 16}, 1519 (1977);
  R.~Casalbuoni, D.~Dominici, R.~Gatto and C.~Giunti,
  Phys.\ Lett.\ B {\bf 178}, 235 (1986);
  R.~Casalbuoni, D.~Dominici, F.~Feruglio and R.~Gatto,
  Nucl.\ Phys.\ B {\bf 299}, 117 (1988).



\bibitem{unitarity2}
  H.~Huffel and G.~Pocsik,
  Z.\ Phys.\ C {\bf 8}, 13 (1981);
  J.~Maalampi, J.~Sirkka and I.~Vilja,
  Phys.\ Lett.\ B {\bf 265}, 371 (1991);
  S.~Kanemura, T.~Kubota and E.~Takasugi,
  Phys.\ Lett.\ B {\bf 313}, 155 (1993)
  [hep-ph/9303263];
  A.~G.~Akeroyd, A.~Arhrib and E.~-M.~Naimi,
  Phys.\ Lett.\ B {\bf 490}, 119 (2000)
  [hep-ph/0006035].




\bibitem{unitarity2HDM}
  I.~F.~Ginzburg and I.~P.~Ivanov,
  Phys.\ Rev.\ D {\bf 72}, 115010 (2005)
  [hep-ph/0508020];
  J.~Horejsi and M.~Kladiva,
  Eur.\ Phys.\ J.\ C {\bf 46}, 81 (2006)
  [hep-ph/0510154].


\bibitem{Altarelli:1990zd}
  G.~Altarelli and R.~Barbieri,
  Phys.\ Lett.\ B {\bf 253}, 161 (1991).

\bibitem{STUcurrent}
  M.~Baak, M.~Goebel, J.~Haller, A.~Hoecker, D.~Kennedy, R.~Kogler, K.~Moenig and M.~Schott {\it et al.},
  Eur.\ Phys.\ J.\ C {\bf 72}, 2205 (2012)
  [arXiv:1209.2716 [hep-ph]].




\bibitem{hgmgmidm1}
A.~Djouadi,
  Phys.\ Rept.\  {\bf 457} (2008) 1
  [arXiv:hep-ph/0503172];
A.~Djouadi,
 Phys. Rept., 459, 2008, pages 1-241,
  arXiv:hep-ph/0503173.




\bibitem{hgmgmidm3}
  A.~Arhrib, R.~Benbrik and N.~Gaur,
  Phys.\ Rev.\ D {\bf 85}, 095021 (2012)
  [arXiv:1201.2644 [hep-ph]].


\bibitem{hgmgmidm4}
  B.~Swiezewska and M.~Krawczyk,
  arXiv:1212.4100 [hep-ph].



\bibitem{SUSEFLAV}
  D.~Chowdhury, R.~Garani and S.~K.~Vempati,
  Comput.\ Phys.\ Commun.\  {\bf 184}, 899 (2013)
  [arXiv:1109.3551 [hep-ph]].



\bibitem{STIDM}
  R.~Barbieri, L.~J.~Hall and V.~S.~Rychkov,
  Phys.\ Rev.\ D {\bf 74}, 015007 (2006)
  [hep-ph/0603188].




\end{thebibliography}
\end{document}